\documentclass[10pt,journal,compsoc]{IEEEtran}

\usepackage[utf8]{inputenc}
\usepackage[linesnumbered]{algorithm2e}
\usepackage[utf8]{inputenc}
\usepackage{microtype}
\usepackage{booktabs}
\usepackage{flushend}
\usepackage{hyperref}
\usepackage[inline]{enumitem}
\usepackage[numbers]{natbib}

\newcounter{observations}
\newcommand{\observation}[1]{\refstepcounter{observations} \vspace{2mm} \noindent \textbf{Observation \theobservations: #1}}

\usepackage[most]{tcolorbox}

\newcommand{\ssubsection}[1]{\vspace{1mm} \noindent \textbf{#1}.}

\begin{document}

\title{The Effectiveness of Supervised Machine Learning Algorithms in\\Predicting Software Refactoring}

\author{\IEEEauthorblockN{Maurício Aniche\IEEEauthorrefmark{1}, Erick Maziero\IEEEauthorrefmark{2}, Rafael Durelli\IEEEauthorrefmark{2}, Vinicius H. S. Durelli\IEEEauthorrefmark{3}\\}
	\IEEEauthorblockA{\IEEEauthorrefmark{1}Delft University of Technology - The Netherlands\\}
	\IEEEauthorblockA{\IEEEauthorrefmark{2}Federal University of Lavras - Brazil\\}
	\IEEEauthorblockA{\IEEEauthorrefmark{3}Federal University of São João del Rei - Brazil\\}
	\IEEEauthorblockA{m.f.aniche@tudelft.nl, erick.maziero@ufla.br, rafael.durelli@ufla.br, durelli@ufsj.edu.br}}

\IEEEtitleabstractindextext{
\begin{abstract}

Refactoring is the process of changing the internal structure of software to improve its quality 
without modifying its external behavior.
Empirical studies have repeatedly shown that refactoring
has a positive impact on the understandability and maintainability of software systems.
However, 
before carrying out refactoring activities, 
developers need to identify refactoring opportunities. 
Currently, 
refactoring opportunity identification heavily relies on developers' expertise and intuition.
In this paper, we investigate the effectiveness of machine learning algorithms 
in predicting software refactorings.
More specifically, 
we train six different machine learning algorithms 
(i.e., Logistic Regression, Naive Bayes, Support Vector Machine, Decision Trees, 
Random Forest, and Neural Network) 
with a dataset comprising over two million refactorings from
11,149 real-world projects from the Apache, F-Droid, and GitHub ecosystems. 
The resulting models predict 20 different refactorings at class, method, and variable-levels 
with an accuracy often higher than 90\%.
Our results show that (i) Random Forests are the best models for predicting software refactoring, 
(ii) process and ownership metrics seem to play a crucial role in the creation of better models, 
and
(iii) models generalize well in different contexts.

\end{abstract}

}

\maketitle

\IEEEdisplaynontitleabstractindextext
\IEEEpeerreviewmaketitle

\setlist[itemize]{leftmargin=*}

\section{Introduction}
\label{sec:intro}

\IEEEPARstart{R}{efactoring}, as defined by Fowler~\cite{Fowler1999-go}
is 
``\textit{the process of changing a software system in such a way that does not alter the external behavior of the code yet improves its internal structure}''. 
Over the years, empirical studies have established a positive correlation between refactoring operations and code quality metrics (e.g., \cite{Eman_2019}, \cite{Kataoka_2012}, \cite{leitch2004assessing}, \cite{alshayeb2009empirical}, \cite{shatnawi2011empirical}).
All these evidence
indicates that refactoring should be regarded as a first-class concern of software developers. 

However, \textbf{deciding when and what (as well as understanding why) to refactor} have long posed a challenge 
to developers.
Software development teams should not simply refactor their software systems at will, or decide not to refactor a piece of code that causes technical debt, as 
any refactoring activity comes with costs~\cite{kim2012field,kruchten2012technical}.

To that aim, software developers have been relying more and more on different static analysis tools and linters as a way to collect feedback about their source code~\cite{beller2016analyzing}. Developers not only use these tools to find bug-related issues in their systems (e.g., ~\cite{ayewah2008using,habchi2018adopting}), but also for code quality-related advice ~\cite{tomasdottir2018adoption, tomasdottir2017and}. Popular tools such as PMD, ESLint, and Sonarqube offer detection strategies for common code smells, such as 
\emph{God Classes} or \emph{Long Methods}. 
These tools have been now integrated into different stages of the developers' workflow, e.g., inside IDEs (e.g., PMD's plugin for IntelliJ or Eclipse), during code review (by means of bots), or as a overall quality report (for example, Sonarqube's Technical Debt report).

Identifying refactoring opportunities is an important stage that precedes the refactoring process. However, despite their importance to the software development world, the state-of-the-art tools that developers have been using to get refactoring recommendations often present a high number of false positives~\cite{johnson2013don}, making developers to lose their confidence on them. The tools' detection strategies are often either based on hard thresholds of single metrics (e.g., PMD considers all methods with more than 100 lines of code, ``problematic''), or on Lanza's and Marinescu's seminal work on code smells detection strategies~\cite{lanza2007object} which rely on a combination of code metrics and thresholds. 

While tools provide some degree of customization, e.g., PMD lets developers choose their own thresholds, and Decor~\cite{moha2009decor} enables developers to devise their own code smells detection strategies, such hand-made detection strategies may be too simplistic to capture the full complexity of software systems. This is where we conjecture a ML-based solution would help. \textbf{We argue that the task of identifying relevant refactoring opportunities,
which currently heavily relies on developers' expertise and intuition,
should be supported by sophisticated recommendation algorithms.}

Researchers have been indeed experimenting with different AI-based techniques to recommend
refactoring, 
such as the use of search algorithms~\cite{mariani2017systematic,o2008search}, and pattern mining~\cite{bavota2014recommending}.
In this paper, 
we explore how machine learning (ML) can be harnessed to predict refactoring operations. 
ML algorithms have been showing promising results when applied to different
areas of software engineering, 
such as defect prediction~\cite{d2012evaluating}, code comprehension~\cite{Liu_2019}, and code smells~\cite{Muhammad_2019}.
By learning from classes and methods that underwent refactoring operations in practice, we surmised that the resulting models would be able to provide more reliable refactoring recommendations to developers.

We formulate the prediction of refactoring opportunities as a binary classification problem. 
We build models that recommend several different refactoring operations (the full list of refactoring operations is shown in Table~\ref{tab:refactorings}). 
Each model predicts whether a given piece of code should undergo a specific refactoring operation. 
For instance, 
given a method, the \emph{Extract Method} model predicts whether that method should undergo a extract method refactoring operation. 
More formally, given
a set $R$ of possible refactorings for a source code element, 
we learn a set of models $M_r(e)$, $r~\exists~R$, 
that predict whether a source code element $e$ should be refactored by means of refactoring operation $r$.

To probe into the effectiveness of supervised ML algorithms 
in predicting refactoring opportunities, 
we apply six ML algorithms
(i.e., Logistic Regression, Naive Bayes, Support Vector Machines, Decision Trees, Random Forest, and
Neural Network) to a dataset containing more than
two million labelled refactoring operations that happened in 11,149 open-source projects 
from the Apache, F-Droid, and GitHub ecosystems.
The resulting models are able to predict
20 different refactoring operations at class, method, and variable-levels~\cite{Fowler1999-go}, with an
average accuracy often higher than 90\%.

Understanding the effectiveness of the different models is the \textbf{first and necessary step in building tools that will help developers in drawing data-informed refactoring decisions}.
This paper provides the \textbf{first solid large-scale evidence that ML algorithms can model the refactoring recommendation problem accurately}. 

In summary, 
this paper makes the following contributions:

\begin{enumerate}[label=(\roman*), font=\itshape]
  \item A large-scale in-depth study of the effectiveness of different
  supervised ML algorithms to predict software refactoring, showing that ML methods can accurately model the refactoring recommendation
  problem.
  \item A dataset containing more than two million real-world 
  refactorings extracted from more than 11 thousand real-world projects.
\end{enumerate}

\section{Research Methodology}

\begin{figure*}
 \centering 
 \includegraphics[width=0.97\textwidth]{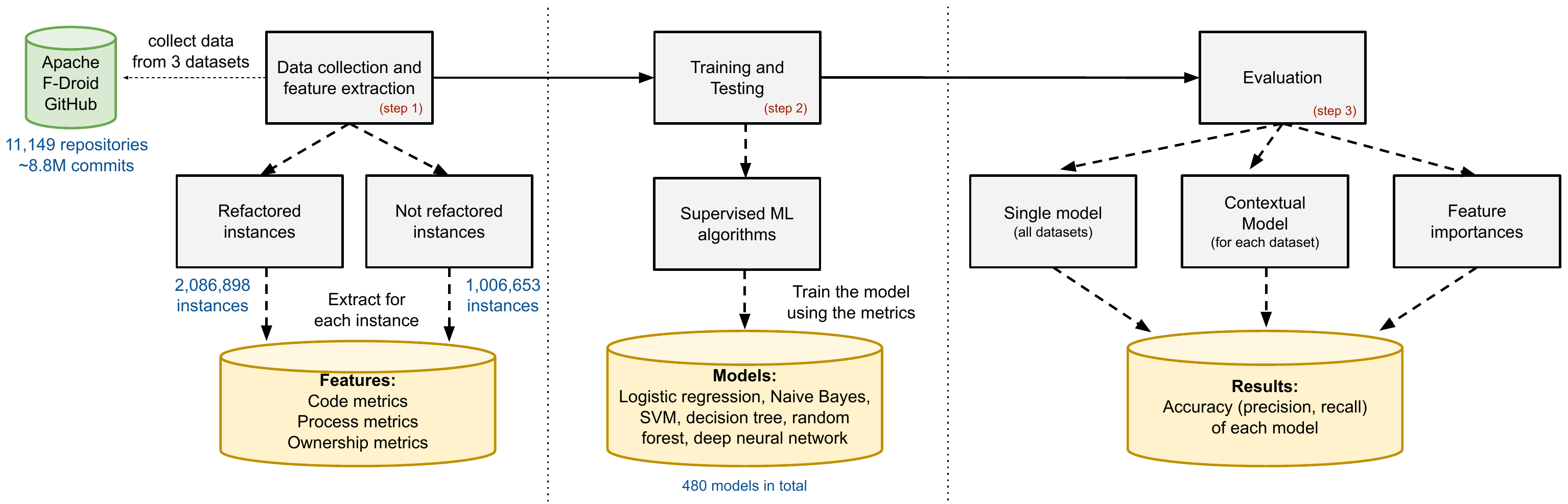}
 \caption{Overview of the research methodology.}
 \label{fig:methodology}
\end{figure*}

The goal of this paper is to
\textbf{evaluate the feasibility of using supervised ML algorithms to identify refactoring opportunities}. 
To this end, 
we framed our research around the following research questions (RQs):

\newcommand{\rqone}{How accurate are supervised ML algorithms in predicting software refactoring?}

\newcommand{\rqtwo}{What are the important features in the refactoring prediction models?}

\newcommand{\rqthree}{Can the predictive models be carried over to different contexts?}

\smallskip
\textbf{RQ$_1$: \rqone}
In practice, some prediction algorithms perform better than others, depending on the task.
In this RQ, 
we explore how accurate different supervised ML algorithms (i.e., Support Vector Machines, Naive Bayes, Decision Trees, Random Forest, and Neural networks)
are in predicting refactoring opportunities at different levels (i.e., refactorings at class, method, and variable-levels), 
using Logistic Regression as a baseline for comparison. 
    
\smallskip
\textbf{RQ$_2$: \rqtwo}
Features (i.e., 
a numeric representation of a measurable property that is used to represent a ML problem to the model) play a pivotal role in the quality of the obtained models.
In RQ$_1$, we build the models using all the features we had available (for a method-level refactoring,
for example, we use 58 different features).
In this RQ, 
we explore which features are considered the most relevant
by the models. 
Such knowledge is essential because, in practice,
models should be as simple as and require as little data as possible.
    
\smallskip
\textbf{RQ$_3$: \rqthree}
Understanding whether refactoring prediction models should be trained specifically for a given context
or whether it generalizes enough to different contexts
can significantly reduce the
cost of applying and re-training these models in practice. 
We set out to study whether prediction models, devised in one type of software systems (e.g., libraries and frameworks from the Apache ecosystem), 
are able to generalize to 
different types of software systems (e.g., mobile apps in the F-Droid ecosystem).
We investigate the accuracy of predictive models against independent datasets (i.e., out-of-sample accuracy). 

\smallskip

Figure~\ref{fig:methodology} shows an overview of the approach we used to answer the aforementioned RQs. 
Essentially, our approach is three-fold:
\begin{enumerate*}[label=(\roman*), font=\itshape]
\item data collection and feature extraction, 
\item training and testing, and 
\item evaluation. 
\end{enumerate*}
These steps are outlined below and later better detailed in the following subsections.

The first step is centered around data preparation.
This step involves mining software repositories for labelled instances of \textit{refactored elements}, e.g., a method that was moved,
or an inlined variable, and instances of \textit{elements that were not refactored}. 
To both refactored and non-refactored instances,
we extract code metrics (e.g., complexity and coupling), process metrics (e.g., number of commits in that class), and ownership metrics (e.g., number of authors). The code metrics are calculated at different levels, depending on the type of refactoring. For a class-level refactoring, we calculate class-level metrics; for a method-level refactoring, we calculate both class and method-level metrics; for a variable-level refactoring, we calculate class, method, and variable-level metrics. 

In the second step, we use the examples of refactored and non-refactored elements we collected as training and testing data to different ML algorithms.
We generate a model for each combination of datasets (all datasets together, Apache, F-Droid, and GitHub),
refactoring operations (the 20 refactoring operations we show in Table~\ref{tab:refactorings}), 
and ML algorithms (i.e., Logistic Regression, Naive Bayes, Support Vector Machines, Decision Tree, Random Forest, and Neural Network). 
Before training  
the final model, our pipeline balances the dataset, performs a random search for 
the best hyper-parameters, and stores the best configuration and 
the ranking of importance of each feature.

In the third step, we evaluate the accuracy of each generated model. First, 
we test the model using single datasets. Next, we test the models that were 
trained using data from just one dataset and test it in all the other datasets 
(e.g., the model trained with the Apache dataset is tested on the GitHub and F-Droid datasets). 
In all the runs, we record the model's precision, recall, and accuracy. 

\begin{table}
\small
\centering
\begin{tabular}{lrr}
\toprule
& Number of  & Total number \\
& Projects  &  of commits \\
\midrule
Apache          & 844                      & 1,471,203 \\
F-Droid         & 1,233                          & 814,418   \\
GitHub          & 9,072                       & 6,517,597 \\
\midrule
                & \textbf{11,149}                   & \textbf{8,803,218}\\
\bottomrule
\end{tabular}
\caption{Overview of the sample used in our study.}
\label{tab:sample}
\vspace{-5mm}
\end{table}

\subsection{Experimental Sample}
\label{sec:sample}

\begin{table*}
\begin{tabular}{lp{13cm}}
\toprule
\textbf{Refactoring}                     & \textbf{Problem and Solution}                                                                                                                                                                                                                                                \\ 
\midrule
\multicolumn{2}{l}{\textbf{Class-level refactorings}}    \\  
\hspace{2mm}Extract Class                   & A class performs the work of two or more classes. Create a class and move the fields and methods to it. \\
\hspace{2mm}Extract Subclass                & A class owns features that are used only in certain scenarios.  Create a subclass.                                                      \\
\hspace{2mm}Extract Super-class              & Two classes own common fields and methods. Create a super class and move the fields and methods.                           \\
\hspace{2mm}Extract Interface               & A set of clients use the same part of a class interface. Move the shared part to its own interface.                                                                                       \\
\hspace{2mm}Move Class                      & A class is in a package with non-related classes.  Move the class to a more relevant package.                  \\
\hspace{2mm}Rename Class                    & The class' name is not expressive enough.  Rename the class.                                                                                                                  \\
\hspace{2mm}Move and Rename Class           & The two aforementioned refactorings together. \\ 
\midrule
\multicolumn{2}{l}{\textbf{Method-level refactorings}}        \\ 

\hspace{2mm}Extract Method                  & Related statements that can be grouped together.  Extract them to a new method.                          \\
\hspace{2mm}Inline Method                   & Statements unnecessarily inside a method.  Replace any calls to the method with the method's content.                                                   \\
\hspace{2mm}Move Method                     & A method does not belong to that class. Move the method to its rightful place.               \\
\hspace{2mm}Pull Up Method                  & Sub-classes have methods that perform similar work. Move them to the super class.                                                                   \\
\hspace{2mm}Push Down Method                & The behavior of a super-class is used in few sub-classes. Move it to the sub-classes.                                                                                              \\
\hspace{2mm}Rename Method                   & The name of a method does not explain the method's purpose.  Rename the method.                                                                                                                 \\ 
\hspace{2mm}Extract And Move Method         & The two aforementioned refactorings together. \\
\midrule                                  
\multicolumn{2}{l}{\textbf{Variable-level refactorings}}     \\ 

\hspace{2mm}Extract Variable                & Hard-to-understand/long expression. Divide the expression into separate variables.                                    \\
\hspace{2mm}Inline Variable                 & Non-necessary variable holding an expression. Replace the variable references with the expression itself.                                                                 \\
\hspace{2mm}Parameterize Variable           &  Variable should be a parameter of the method. Transform variable into a method parameter. \\
\hspace{2mm}Rename Parameter                & The name of a method parameter does not explain its purpose. Rename the parameter.                                                                                                               \\
\hspace{2mm}Rename Variable                 & The name of a variable does not explain the variable's purpose. Rename the variable.                                                                                                               \\
\hspace{2mm}Replace Variable w/ Attribute &   Variable is used in more than a single method. Transform the variable to a class attribute.      \\ 
\bottomrule                            
\end{tabular}
\caption{The 20 refactoring operations that are studied in this paper. Definitions derived from Fowler~\cite{Fowler1999-go}.}
\label{tab:refactorings}
\vspace{-5mm}
\end{table*}

We selected a very large and representative set of Java projects from three different sources: 

\begin{itemize}
    \item The Apache Software Foundation (ASF) is a non-profit organization
    that supports all Apache software projects. The ASF is responsible for projects such as Tomcat, Maven, and Ant.
    Our tools successfully processed 844 out of their 860 Java-based projects. 
    We discuss why the processing of some projects have failed in Section~\ref{sec:execution}.
    
    \item F-Droid is a software repository of Android mobile apps. The repository contains only free software apps. Our tools successfully processed 1,233 out of their 1,352 projects.
    
    \item GitHub provides free hosting for open source projects. GitHub has been extensively used by the open source community. As of May 2019, GitHub has 37 million users registered. We collected the first 10,000 most starred Java projects. Note that ASF and F-Droid projects might also exist in GitHub; we removed duplicates. In the end, our tools were able to process 9,072 projects. 
\end{itemize}

The three different sources of projects provide the dataset with
high variability in terms of size and complexity
of projects, domains and technologies used, and community. 
The resulting sample can be seen in Table~\ref{tab:sample}. It 
comprises the 11,149 projects (844 from Apache, 1,233 from F-Droid, and 9,072 from GitHub).
These projects together 
a history of 8.8 million commits, measured at the 
moment of data collection, in March of 2019.

\subsection{Extraction of Labelled Instances}

In a nutshell, our data collection process happens in three phases. In the first phase,
the tool clones the software repository, uses RefactoringMiner~\cite{Tsantalis_2018} to collect refactoring operations
that happened throughout the history of the repository, and collects the code metrics of the
refactored classes. In the second phase, where all the refactoring operations and their respective files are already known, the tool then collects the process and ownership metrics of the refactored classes. Finally,
the tool  
collects instances
of non-refactored classes (as well as their code, process, and ownership metrics).

For each project, we visit its entire master branch from the oldest to the most recent commit. For each commit, we invoke RefactoringMiner~\cite{Tsantalis_2018}. The tool can receive, as an input, a pair of commits. It then uses the \textit{diff} between the two provided commits to identify refactoring operations that have happened\footnote{Given that we need a pair of commits in order to identify the refactoring operations, we skip the first commit of the repository, and start from commit no. 2.}. We highlight that RefactoringMiner is the current state-of-the-art tool to identify refactoring operations, having the highest recall and precision rates (98\% and 87\%, respectively) among all currently available refactoring detection tools~\cite{Tsantalis_2018}. 

For each refactoring operation that is detected by RefactoringMiner, we extract code metrics of the refactored element in its version \textit{before} the refactoring has been applied. The intuition behind using the version before the refactoring is that models should learn how to identify refactorings by looking at the elements as they were prior to being refactored. We collect the information at the precise level of the refactoring. For example, if the refactoring is at class-level, we collect all the class-level metrics related to the class under refactoring; if it is a method-level refactoring, we collect metric-level metrics related to the method under refactoring; the same applies for variable-level refactorings.

After all the refactorings were identified, our tool collects the
process and ownership metrics of the refactored classes. These metrics are also collected at
the version \textit{before} the refactoring had been applied. 

Finally, our tool collects instances of \textit{non-refactored
classes, methods, and variables}, i.e., code elements that
did not undergo any refactoring operations, to serve as counterexamples 
to the model. This is a fundamental step as 
binary classification models should learn how to separate 
between the two classes; in this case, between
methods that need to be refactored, and methods that do not need to be refactored.

Given that there is no clear way of extracting code elements that do not
need to be refactored out of the source code history of software systems, 
we propose an heuristic:
we consider a class to be a non-refactoring instance if it was modified (i.e., a change committed in the Git repository) precisely $k$ times without a single refactoring operation being applied in between this time.
The heuristic aims at identifying classes that can still be evolved by developers (as developers have been evolving them) without the need for a refactoring (as we see that they did not apply any refactorings).
We conjecture that such classes can serve as good counterexamples for
the model.

After experimentation, we set $k = 50$ (we discuss the influence of $k$ 
in Section~\ref{sec:non-refactored-instances}). The tool, therefore, collects all classes that were modified precisely $k$ times and did not go through any refactoring operation. We then extract its source code, process, and ownership metrics. Note that we extract the metrics at time $0$, 
and not at time $k$, as we want the models to learn from the code element that, back then, did not require any refactoring from developers. The same element can appear more than once in this dataset (although always with different metric values), as whenever we collect an instance of non-refactoring, we restart its counter and continue to visit the repository.

Note that our approach ignores test code (e.g., JUnit files) and only captures refactoring operations in production files. Test code quality has been the target of many studies (e.g.,~\cite{van2001refactoring,bavota2012empirical,bavota2015test}). In this work, we assume that refactorings that happen in test code are naturally different from the ones that happen in production code; our future agenda includes
the development of refactoring models for test code.

In Table~\ref{tab:sample-refactorings}, we show the number of refactored and non-refactored instances we collected per dataset. 
We highlight the fact that the number of instances varies per refactoring, 
which reflects how much developers apply each of these refactorings. 
For example, the dataset contains around 327 thousand instances of \textit{Extract Method}, but only 654 instances of \textit{Move and Rename Class}. 
We see this as a positive point to our exploration, as the model will have to deal with refactorings where the number of instances is not high. 

\begin{table}
\begin{tabular}{p{3cm}rrrr}
\toprule
& All           & Apache & GitHub & F-Droid  \\
\midrule
\multicolumn{5}{l}{\textbf{Class-level refactorings}} \\
\hspace{2mm}Extract Class                   & 41,191  & 6,658   & 31,729   & 2,804   \\
\hspace{2mm}Extract Interface               & 10,495  & 2,363   & 7,775    & 357    \\
\hspace{2mm}Extract Subclass                & 6,436   & 1,302   & 4,929    & 205    \\
\hspace{2mm}Extract Superclass              & 26,814  & 5,228   & 20,027   & 1,559   \\
\hspace{2mm}Move And Rename Class           & 654    & 87     & 545     & 22     \\
\hspace{2mm}Move Class                      & 49,815  & 16,413  & 32,259   & 1,143   \\
\hspace{2mm}Rename Class                    & 3,991   & 557    & 3,287    & 147    \\
\midrule
\multicolumn{5}{l}{\textbf{Method-level refactorings}}\\
\hspace{2mm}Extract And Move Method         & 9,723   & 1,816   & 7,273    & 634    \\
\hspace{2mm}Extract Method                  & 327,493 & 61,280  & 243,011  & 23,202  \\
\hspace{2mm}Inline Method                   & 53,827  & 10,027  & 40,087   & 3,713   \\
\hspace{2mm}Move Method                     & 163,078 & 26,592  & 124,411  & 12,075  \\
\hspace{2mm}Pull Up Method                  & 155,076 & 32,646  & 116,953  & 5,477   \\
\hspace{2mm}Push Down Method                & 62,630  & 12,933  & 47,767   & 1,930   \\
\hspace{2mm}Rename Method                   & 427,935 & 65,667  & 340,304  & 21,964  \\
\midrule
\multicolumn{5}{l}{\textbf{Variable-level refactorings}}\\
\hspace{2mm}Extract Variable                & 6,709   & 1,587   & 4,744    & 378    \\
\hspace{2mm}Inline Variable                 & 30,894  & 5,616   & 23,126   & 2,152   \\
\hspace{2mm}Parameterize Variable           & 22,537  & 4,640   & 16,542   & 1,355   \\
\hspace{2mm}Rename Parameter                & 33,6751 & 61,246  & 261,186  & 14,319  \\
\hspace{2mm}Rename Variable                 & 324,955 & 57,086  & 250,076  & 17,793  \\
\hspace{2mm}Replace Variable w/ Attr. 
 & 25,894      & 3,674   & 18,224   & 3,996   \\
\midrule
\multicolumn{5}{l}{\textbf{Non-refactoring instances}}\\
\hspace{2mm}Class-level         & 10,692      & 1,189   & 8,043    & 1,460   \\
\hspace{2mm}Method-level        & 293,467 & 38,708  & 236,060  & 18,699  \\
\hspace{2mm}Variable-level      & 702,494 & 136,010 & 47,811   & 518,673 \\
\bottomrule
\end{tabular}
\caption{Overview of the number of instances of refactoring and non-refactoring classes.}
\label{tab:sample-refactorings}
\end{table}

\subsection{Feature Selection}
\label{sub:feature_extraction}

\begin{table}
\begin{tabular}{p{0.93\columnwidth}}
\toprule
\textbf{Class-level (total of 46 metrics)} \\
\textbf{Source Code (37 metrics):}
CBO, WMC, RFC, LCOM, number of methods, number of static methods, number of public methods, number of private method, number of protected method, number of abstract methods, number of final methods, number of synchronized methods, number of fields, number of static fields, number of public fields, number of private fields, number of protected fields, number of default fields, number of final fields, number of synchronized fields, number of static invocations, lines of code, number of 'return' statements, number of loops, number of comparison expressions, number of try catches, number of expressions with parenthesis, number of string literals, number of 'number constants', number of assignments, number of mathematical operators, number of declared variables, max number of nested blocks, number of anonymous classes, number of sub classes, number of lambda expressions, number of unique words. \\

\textbf{Process (5 metrics):} Quantity of commits, sum of lines added, sum of lines deleted, number of bug fixes, number of previous refactoring operations. \\
\textbf{Ownership (4 metrics):} Quantify of authors, quantity of minor authors, quantity of major authors, author ownership. \\
\midrule 
\textbf{Method-level (total of 20 metrics + 37 code metrics at class-level)} \\
Source Code (20 metrics):
CBO, WMC, RFC, lines of code, number of 'return' statements, number of variables, number of parameters, number of loops, number of comparison operators, number of try/catches, number of expressions with parenthesis, number of string literals, number of 'number constants', number of assignment, number of mathematical operators, max number of nested blocks, number of anonymous classes, number of sub-classes, number of lambda expressions, number of unique words. \\
\midrule
\textbf{Variable-level (total of 1 metric + 57 method+class level)} \\
Source Code (1 metric): Number of times the variable is used. \\
\bottomrule
\end{tabular}
\caption{List of features collected at class, method, and variable levels.}
\label{tab:features}
\vspace{-8mm}
\end{table}

We extract source code, process, and ownership metrics of all refactored and non-refactored instances. 
These three types of metrics have been proven useful in other prediction models in software
engineering (e.g.,~\cite{scalabrino2017automatically,mcintosh2014impact,d2012evaluating}).
Moreover, 
earlier studies based on the correlation between refactoring and code quality metrics postulated that an increase in the former 
leads to improvements in the latter
(e.g., \cite{Kataoka_2012}, \cite{leitch2004assessing}, \cite{shatnawi2011empirical}).\footnote{It is worth noting that studying the effect of refactoring on software quality is a topic that remains relatively underdeveloped (despite being a highly active topic). 
Therefore, while this research topic is evolving, the evidence is likely to be far from clear-cut and, in some cases, it might even be contradictory (e.g., \cite{how_does_oo_code_refactoring_influence_software_quality,alshayeb2009empirical}).}
Table~\ref{tab:features} lists all the metrics we chose to train predictive models. In our online appendix~\cite{appendix}, we
show the distribution (i.e., descriptive statistics) of the values of each feature.
The following subsections detail the source code, process, and code ownership metrics we collect. 

\ssubsection{Source Code Metrics} 
Features in this category are derived from source code attributes.  
We collect CK metrics~\cite{chidamber1994metrics} as they express the complexity of the element. 
More specifically, 
CBO, WMC, RFC, and LCOM. We also collect several different attributes of the element, e.g., number of fields, number of loops, number of return statements.
These metrics are collected at class (37 metrics), method (20 metrics), and variable-levels (1 metric). 

\ssubsection{Process Metrics} 
Process metrics have been proven useful in defect prediction algorithms~\cite{rahman2013and,madeyski2015process}.
We collect five different process metrics: quantity of commits, the sum of lines added and removed, number of bug fixes, and number of previous refactoring operations.
The number of bug fixes is calculated by means of an heuristic: Whenever any of the keywords \{\texttt{bug}, \texttt{error}, \texttt{mistake}, \texttt{fault}, \texttt{wrong}, \texttt{fail}, \texttt{fix}\} appear in the commit message, we count one more bug fix to that class. The number of previous refactoring operations is based on the refactorings we collect from RefactoringMiner.

\ssubsection{Code Ownership Metrics}
We adopt the suite of ownership metrics proposed by Bird et al.~\cite{dont_touch_my_code}. 
The \emph{quantity of authors} is the total number of developers that have contributed to the given software artifact. 
The \emph{minor authors} represent the number of
contributors that authored less than 5\% (in terms of the number of commits) of an artifact.
The \emph{major authors} represent the number of developers that
contributed at least 5\% to an artifact. Finally,
\emph{author ownership} is the proportion of commits achieved by the most active developer.

The cardinality of the set of features we use to train each model varies. 
The feature set for training models whose desired output is to predict class-level refactoring 
comprises 46 features: 
37 source code metrics, 
5 process metrics, and 4 ownership metrics.
As for the training of method-level models, 
we use a set of features that comprises all the 37 class-level source code metrics 
plus 20 method-level source code metric features, 
totaling 57 features. 
The same holds for variable-level models, 
all class, method, and variable-level source code metrics features are used to fit these models.

Process and ownership metrics are only used in class-level refactoring models. Our tool relies on Git data to measure ownership and process metrics. However, Git provides information solely at file and line levels. While process and ownership metric values for a file are good approximations of process and ownership metric values for classes, the same does not hold for methods and variables. Technically speaking, extracting such metrics in a fine-grained manner (i.e., which methods or variables were modified, precisely) would cost extra computational analysis, which we decided to avoid. We discuss the importance of such metrics later in Section~\ref{subsec:importance-of-other-metrics}.\footnote{Recent work by Higo et al.\cite{higo2020tracking} proposes a ``finer Git'', which tracks changes in individual methods. Such tool was not available at the time of this research.

By using class-level features in the training of method-level refactoring prediction models (or similarly, class-level and method-level features in variable-level refactoring models), we give models a ``sense of context''. The intuition is that developers might not decide to refactor a method by only looking at it; rather, they might look at the overall context (i.e., class) that the method belongs to.

Subsequently, the input to a trained model is a feature vector containing the source code, process, and ownership metrics of the class, method, or variable one wants to predict.
}

\subsection{Model Training}

In this step, we train
different ML algorithms to predict refactoring opportunities. 
We use the collected refactoring instances (and their non-refactoring counterexamples) as training data.

We make use of six different (binary classification) supervised
ML algorithms,
all available in the scikit-learn~\cite{scikit-learn} and keras: 

\begin{enumerate}[label=(\roman*), font=\itshape]
\item Logistic Regression~\cite{bishop2006_lr}: 
Logistic Regression is, similarly to linear regression, centered on combining input values using coefficient values (i.e., weights) to predict an outcome value.  However, differently from linear regression, the outcome value being modeled ranges from 0 to 1. 

\item (Gaussian) Naive Bayes \cite{Zhang04theoptimality}:
Naive Bayes algorithms describe a set of steps to apply Bayes' theorem to classification problems.  
These algorithms use training data to compute the probability of each outcome based on the information extracted from feature values.

\item Support Vector Machines~\cite{cortes1995_svm}: 
Support Vector Machines computes a hyper-plane 
in a high-dimensional space to classify data into predefined classes. 
The algorithm searches for the best hyper-plane to separate the training instances into their 
respective classes. 

\item Decision Trees~\cite{quinlan1993_dt}: 
Decision Tree algorithms yield hierarchical models composed of decision nodes and leaves. 
Essentially, the resulting models represent a partition of the feature space. 

\item Random Forest~\cite{breiman2001_rf}:
Random Forest is an ensemble of decision tree predictors.  
That is, such algorithm uses a number of decision trees with random subsets of the training data.

\item Neural Networks \cite{Goodfellow-et-al-2016}: 
Neural Networks are a family algorithms designed to loosely resemble how the human brain processes information. 
The elements that comprise the architecture of such algorithms are similar to neurons, 
and Neural Networks are made up of one or more layers of these neurons. Essentially, 
these layers of neurons act as a function, mapping inputs into their respective classes. 

\end{enumerate}

We decided to choose a mixture of simple/less sophisticated learners (e.g., Logistic Regression and Naive Bayes) and smarter learners (e.g., Decision Trees and Random Forests). Simple learners serve as a baseline to understand whether more complex learners are needed.

Our training pipeline works as follows:

\begin{enumerate}[leftmargin=*, label=(\roman*),  labelwidth=!, labelindent=0pt, font=\itshape]
    \item We collect the refactoring 
    and the non-refactoring instances for a given dataset $d$ and a refactoring $R$. We merge them in a single dataset, where refactoring instances are marked with a \textit{true} value and non-refactoring instances are marked with a \textit{false} value. These instances will later serve as training and test data.
    
    \item The number of refactoring instances vary per refactoring; thus, the number of refactoring instances might be greater than or smaller than the (fixed) number of non-refactoring instances. Thus, we balance the dataset as to avoid the model to favour the majority class. To that aim, we use scikit-learn's random under sampling algorithm, which randomly selects instances of the over-sampled class.\footnote{We discuss the impact of balancing the classes in Section~\ref{sec:internal-validity}.}.
    
    \item We scale all the features to a $[0,1]$ range to speed up the learning process of the algorithms \cite{Normalization43442}. We use the Min-Max scaler provided by the scikit-learn framework.
    
    \item We tune the hyper parameters of each model by means of a random search. We use the randomized search algorithm provided by the scikit-learn. We set the number of iterations to 100 and the number of cross-fold validations to 10. Thus, we create 1,000 different models before deciding the model's best parameters. 
	    For the Support Vector Machines (SVM) in particular, we use number of iterations as 10 and number of cross-fold validations to 5, given 
    its slow training time (which we discuss more below). 
    For each algorithm, we search the best configuration among the following parameters:
        \begin{itemize}
            
            \item \textbf{Logistic Regression:} \emph{C}: This parameter specifies, inversely, the strength of the regularization. Regularization is a technique that diminishes the chance of overfitting the model.
            
            \item \textbf{Naive Bayes:} \emph{Smoothing}: It specifies the variance of the features to be used during training.
            
            \item \textbf{SVM:} \emph{C}: This parameter informs the SVM optimization algorithm how much it is desired to avoid misclassifying training instances. Like the \emph{C} parameter in the Logistic Regression, it helps in avoiding overfitting. Moreover, given that our goal
            is to also understand which features are important to the model (RQ$_2$), we opt only for the
            linear kernel of the SVM. Future research should explore how non-linear kernels perform.
            
            \item \textbf{Decision tree:}  
            \emph{Max depth}: It specifies the maximum depth of the generated tree. The deeper the tree, more complex the model becomes; \emph{Max features}: It defines the maximum number of features to be inspected during the search for the best split, generating inner nodes; \emph{Min sample split}: It indicates the minimum number of instances needed to split an internal node, supporting the creation of a new rule; \emph{Splitter}: It defines the strategy in choosing the split at each node, varying from ``best to random'' strategies; \emph{Criterion}: It defines the function to measure the quality of a split.
            
            \item \textbf{Random Forest:} 
            The \emph{max depth}, \emph{max features}, \emph{min samples split}, and \emph{criterion} parameters have similar goals as to the ones in the Decision Tree algorithm; \emph{Bootstrap}: It specifies whether all training instances or bootstrap samples are used to build each tree; \emph{Number of estimators}: It indicates the number of trees in the forest.

            \item \textbf{Neural Network:} As we intend to explore sophisticated and more appropriated Deep Learning architectures in the future work (Section~\ref{sec:discussion}), here we compose a sequential network of three dense layers with 128, 64, and 1 units, respectively.  Also, to avoid overfitting, we added dropout layers between sequential dense layers, keeping the \textit{learning} in 80\% of the units in dense layers. The number of epochs was set to 1000. This architecture is similar to a Multilayer Perceptron, in the sense that it is a feedforward deep network.
            
        \end{itemize}
    
    \item Finally, we perform a stratified 10-fold cross-validation (i.e., 9 folds for training and 1 fold for testing) using the hyper parameters established by the search. We return the precision, recall, and accuracy of all the models.
    
\end{enumerate}

Once a binary classification model for a given refactoring $R$ is trained, given a code element $e$ (i.e., a class, method, or a variable), the model would predict true in case $e$ should undergo through a refactoring $R$, or false in case $e$ should not undergo through a refactoring $R$.\footnote{For completeness, in such models, a false positive would mean that the model predicted true for an element $e$ that, in fact, did not undergo a refactoring $R$; a false negative would mean that a model predicted false for an element $e$ that, in fact, did undergo a refactoring $R$.}
 
\subsection{Evaluation} 

To answer RQ$_1$, we report and compare the mean precision, recall, and accuracy among the different 
models after the 10 stratified cross-fold executions\footnote{We kept the 50-50 distribution in all the 10 folds.}. We apply stratified sampling in all the cross-fold
executions to make sure both training and test datasets contain the same amount of positive and negative instances. 
For SVM and the Neural Network, we set the number
of cross-folds to 5.
The SVM and the Neural Network models training and validation processes took 237 and 232 hours, respectively. 
The precision, recall, and accuracy across the five folds of both models were
highly similar, indicating that the models are stable (numbers can be found in our appendix~\cite{appendix}), and thus, we have no reason to believe that the smaller number of cross-fold validations for the SVM and Neural Network affected their results.

For clarity, we revisit what a correct prediction means in this context. We recall that the feature vectors of the positive labels (i.e., elements that underwent some refactoring operations) are represented by the code metrics collected at the commit right before developers refactored them. In other words, the feature vector represents the code element at the moment that the developer decided that it needed to undergo a refactoring. On the other hand, the feature vectors of the negative labels are represented by the code metrics of classes that did not undergo the refactoring operation for $k$ commits in a row. In other words, code that can be maintained for at least 
$k$ commits without undergoing a refactoring. Thus, a correct prediction means that the model was able to predict that a code element with that characteristic underwent a refactoring operation.

For example, let us suppose a method \texttt{m1()} underwent a \emph{Extract Method} in commit 10. 
This means a developer, when working with \texttt{m1()}'s implementation at version 9 decided the method needed a \emph{Extract Method}. 
When testing the model, we give a feature vector representing \texttt{m1()} in commit 9, and we expect the model to return ``true'' (i.e., this method needs a \emph{Extract Method}). 
If the model returned ``false'', that would result in a false negative. 
Moreover, suppose another method \texttt{m2()} that was changed 50 times between commits [10, 200]. 
In none of these changes developers refactored this method. 
When testing the model, we give a feature vector representing method \texttt{m2()} in commit 10, and we expect the model to return ``false'' (i.e., the method does not need a \emph{Extract Method}). 
If the model returned ``true'', that would be an example of a ``false negative''.

To answer RQ$_2$, we report how often each feature (from Table~\ref{tab:features}) appears among the top-1, top-5, and top-10 most important features of all the generated models. We use scikit-learn's ability
to extract the feature importance of the Logistic Regression, SVM, Decision Trees, and Random Forest models.
The framework does not currently have a native way to extract feature importance of Gaussian Naive Bayes and
Neural Networks. We intend to extract the feature importance of both algorithms via ``permutation importance''
in future work. Given the high number of different models we build (we extracted the feature importance of 320 out of the 480 models we created), we have
no reason to believe the lack of these two models would affect the overall findings of this RQ.
Given that the number of features vary per refactoring level, we generate different rankings for
the different levels (i.e., different ranks for class, method, and variable-level refactorings). Some models (e.g., SVM) might return the importance of a feature as a negative number, indicating that the feature is important for the prediction of the negative class. We consider such a feature also important to the overall model, and thus, we build the ranking using the absolute value of feature importance returned by the models. 

Finally, to answer RQ$_3$, we test each of our dataset-specific models on
the other datasets. For example, we test the accuracy of all Apache's models in the GitHub and F-Droid datasets. 
More formally, for each combination of datasets $d_1$ and $d_2$, where $d_1 \neq d_2$, and refactoring $r$ we:
\begin{enumerate*}
\item load the previously trained $r$ model of the $d_1$ dataset,
\item open the data we collected for $r$ of the $d_2$ dataset,
\item apply the same pre-processing steps (i.e., sampling and scaling),
\item use $d_1$'s model to predict all data points of $d_2$'s dataset,
\item and report the precision, recall, and accuracy of the model.
\end{enumerate*}

\subsection{Implementation and Execution}
\label{sec:execution}

The data collection tool is implemented in Java and stores all 
its data in a MySQL database. The
tool integrates natively with RefactoringMiner~\cite{Tsantalis_2018} 
(also written in Java) as well as with the source code metrics
tool. 

The tool gives RefactoringMiner a timeout of 20 seconds per commit to identify a refactoring. We define the timeout as RefactoringMiner performs several operations to identify refactorings, and these operations grow exponentially, according to the size of the commit. Throughout the development of this study, we observed some commits taking hours to be processed. The 20 seconds was an arbitrary number decided after experimentation. In practice, most commits are resolved by the tool in less than a second. Given that its performance is related to the size of the commit and not to the size of the class under refactoring or the number of refactorings in a commit, we do not believe that ignoring commits where RefactoringMiner takes a long time influences our sample in any way.

Given that our tool integrates different tools, there are many opportunities for failures. We have observed
(i) the code metrics tool failing when the class has an invalid structure (and thus, ASTs can not be built), (ii) our tool failing to populate process and ownership metrics of refactored classes (often due to files being moved and renamed multiple times throughout history, which our tool could not track in 100\% of the cases), (iii) 
RefactoringMiner requiring more memory than what is available in the machine. To avoid possible invalid data points, we discard all data points that were involved in any failure (a total of 10\% of the commits we analyzed). 

We had 30 Ubuntu 18.04 LTS (64bits) VMs, each with 1 GB of Ram, 1 CPU core, and 
20 GB of disk available for data collection. These machines, altogether, spent a total
of 933 hours to collect the data. We observe that the majority of projects (around 99\% of them) took less
than one hour to be processed. 159 of them took more than one hour, and 70 of them more than two hours.
A single project took 23 hours.

The ML pipeline was developed in Python. Most of the code relies on the 
\textit{scikit-learn} framework~\cite{scikit-learn} and \textit{keras} for the Neural Networks training.
To the ML training, we had under our disposition two machines:
one Ubuntu 18.04.2 LTS VM, 396GB of RAM, 40 CPU cores, 
and one Ubuntu 18.04.2 LTS VMS with 14 CPUs and 50 GB of RAM. Given the hyperparameter search and cross-validations, our ML pipeline
experimented with a total of 404,080 models.
The overall computation (training and testing) time was approximately 500 hours.

\subsection{Reproducibility} Our online appendix~\cite{appendix} contains:
\begin{enumerate*}[label=(\roman*), font=\itshape]
\item the list of the 11,149 projects analyzed,
\item a spreadsheet with the full results,
\item the source code of the data collection and the ML tools, and
\item a two million refactorings dataset.
\end{enumerate*}

\section{Results}

In the following subsections, 
we answer each of the RQs.

\begin{table*}
\centering
\resizebox{\textwidth}{!}{
\begin{tabular}{lrrr|rrr|rrr|rrr|rrr|rrr}
\toprule
& \multicolumn{3}{c}{\textbf{Logistic}} &  \multicolumn{3}{c}{\textbf{SVM}} & \multicolumn{3}{c}{\textbf{Naive Bayes}} &  \multicolumn{3}{c}{\textbf{Decision}} &  \multicolumn{3}{c}{\textbf{Random}} &  \multicolumn{3}{c}{\textbf{Neural}}    \\
& \multicolumn{3}{c}{\textbf{Regression}}          & \multicolumn{3}{c}{\textbf{(linear)}}    & \multicolumn{3}{c}{\textbf{(gaussian)}}             & \multicolumn{3}{c}{\textbf{tree}}             & \multicolumn{3}{c}{\textbf{Forest}}         & \multicolumn{3}{c}{\textbf{Network}} \\

\midrule

                                & Pr & Re & Acc  & Pr & Re & Acc      & Pr & Re & Acc        & Pr & Re & Acc        & Pr & Re & Acc        & Pr & Re & Acc \\
\midrule

\multicolumn{16}{l}{\textbf{Class-level refactorings}}\\

\hspace{2mm}Extract Class & 0.78 & 0.91 & 0.82 & 0.77 & 0.95 & 0.83 & 0.55 & 0.93 & 0.59 & 0.82 & 0.89 & 0.85 & 0.85 & 0.93 & \textbf{0.89} & 0.80 & 0.94 & 0.85 \\ 
\hspace{2mm}Extract Interface & 0.83 & 0.93 & 0.87 & 0.82 & 0.94 & 0.87 & 0.58 & 0.94 & 0.63 & 0.90 & 0.88 & 0.89 & 0.93 & 0.92 & \textbf{0.92} & 0.88 & 0.90 & 0.89 \\ 
\hspace{2mm}Extract Subclass & 0.85 & 0.94 & 0.89 & 0.84 & 0.95 & 0.88 & 0.59 & 0.95 & 0.64 & 0.88 & 0.92 & 0.90 & 0.92 & 0.94 & \textbf{0.93} & 0.84 & 0.97 & 0.89 \\ 
\hspace{2mm}Extract Superclass & 0.84 & 0.94 & 0.88 & 0.83 & 0.95 & 0.88 & 0.60 & 0.96 & 0.66 & 0.89 & 0.92 & 0.90 & 0.91 & 0.93 & \textbf{0.92} & 0.86 & 0.94 & 0.89 \\ 
\hspace{2mm}Move And Rename Class & 0.89 & 0.93 & 0.91 & 0.88 & 0.95 & 0.91 & 0.69 & 0.94 & 0.76 & 0.92 & 0.95 & 0.94 & 0.95 & 0.95 & \textbf{0.95} & 0.88 & 0.94 & 0.91 \\ 
\hspace{2mm}Move Class & 0.92 & 0.96 & 0.94 & 0.90 & 0.97 & 0.93 & 0.67 & 0.96 & 0.74 & 0.98 & 0.96 & 0.97 & 0.98 & 0.97 & \textbf{0.98} & 0.92 & 0.97 & 0.94 \\ 
\hspace{2mm}Rename Class & 0.87 & 0.94 & 0.90 & 0.86 & 0.96 & 0.90 & 0.63 & 0.96 & 0.69 & 0.94 & 0.91 & 0.93 & 0.95 & 0.94 & \textbf{0.94} & 0.88 & 0.94 & 0.91 \\

\midrule
\multicolumn{16}{l}{\textbf{Method-level refactorings}}\\

\hspace{2mm}Extract And Move Method & 0.72 & 0.86 & 0.77 & 0.71 & 0.89 & 0.76 & 0.63 & 0.94 & 0.69 & 0.85 & 0.75 & 0.81 & 0.90 & 0.81 & \textbf{0.86} & 0.79 & 0.85 & 0.81 \\ 
\hspace{2mm}Extract Method & 0.80 & 0.87 & 0.82 & 0.77 & 0.88 & 0.80 & 0.65 & 0.95 & 0.70 & 0.81 & 0.86 & 0.82 & 0.80 & 0.92 & \textbf{0.84} & 0.84 & 0.84 & \textbf{0.84} \\ 
\hspace{2mm}Inline Method & 0.72 & 0.88 & 0.77 & 0.71 & 0.89 & 0.77 & 0.61 & 0.94 & 0.67 & 0.94 & 0.87 & 0.90 & 0.97 & 0.97 & \textbf{0.97} & 0.77 & 0.85 & 0.80 \\ 
\hspace{2mm}Move Method & 0.72 & 0.87 & 0.76 & 0.71 & 0.89 & 0.76 & 0.63 & 0.93 & 0.70 & 0.98 & 0.87 & 0.93 & 0.99 & 0.98 & \textbf{0.99} & 0.76 & 0.84 & 0.78 \\ 
\hspace{2mm}Pull Up Method & 0.78 & 0.90 & 0.82 & 0.77 & 0.91 & 0.82 & 0.68 & 0.95 & 0.75 & 0.96 & 0.88 & 0.92 & 0.99 & 0.94 & \textbf{0.96} & 0.82 & 0.87 & 0.84 \\ 
\hspace{2mm}Push Down Method & 0.75 & 0.89 & 0.80 & 0.75 & 0.90 & 0.80 & 0.66 & 0.94 & 0.73 & 0.97 & 0.76 & 0.87 & 0.97 & 0.83 & \textbf{0.90} & 0.81 & 0.92 & 0.85 \\ 
\hspace{2mm}Rename Method & 0.77 & 0.89 & 0.80 & 0.76 & 0.90 & 0.80 & 0.65 & 0.95 & 0.71 & 0.78 & 0.84 & 0.80 & 0.79 & 0.85 & \textbf{0.81} & 0.81 & 0.82 & \textbf{0.81} \\

\midrule
\multicolumn{16}{l}{\textbf{Variable-level refactorings}}\\

\hspace{2mm}Extract Variable & 0.80 & 0.83 & 0.82 & 0.80 & 0.83 & 0.82 & 0.62 & 0.94 & 0.68 & 0.82 & 0.83 & 0.82 & 0.90 & 0.83 & \textbf{0.87} & 0.84 & 0.89 & 0.86 \\ 
\hspace{2mm}Inline Variable & 0.76 & 0.86 & 0.79 & 0.75 & 0.87 & 0.79 & 0.60 & 0.94 & 0.66 & 0.91 & 0.85 & 0.88 & 0.94 & 0.96 & \textbf{0.95} & 0.81 & 0.82 & 0.82 \\ 
\hspace{2mm}Parameterize Variable & 0.75 & 0.85 & 0.79 & 0.74 & 0.86 & 0.78 & 0.59 & 0.94 & 0.65 & 0.88 & 0.81 & 0.85 & 0.93 & 0.92 & \textbf{0.92} & 0.80 & 0.83 & 0.81 \\ 
\hspace{2mm}Rename Parameter & 0.79 & 0.88 & 0.83 & 0.80 & 0.88 & 0.83 & 0.65 & 0.95 & 0.71 & 0.99 & 0.92 & 0.95 & 0.99 & 0.99 & \textbf{0.99} & 0.82 & 0.87 & 0.84 \\ 
\hspace{2mm}Rename Variable & 0.77 & 0.85 & 0.80 & 0.76 & 0.86 & 0.79 & 0.58 & 0.92 & 0.63 & 0.99 & 0.93 & 0.96 & 1.00 & 0.99 & \textbf{0.99} & 0.81 & 0.84 & 0.82 \\ 
\hspace{2mm}Replace Variable With Attribute & 0.79 & 0.88 & 0.82 & 0.78 & 0.89 & 0.82 & 0.64 & 0.95 & 0.71 & 0.90 & 0.84 & 0.88 & 0.94 & 0.92 & \textbf{0.93} & 0.79 & 0.92 & 0.84 \\

\bottomrule
\end{tabular}}
\caption{The precision (Pr), recall (Re), and accuracy (Acc) of the different ML models, when trained and tested in the entire dataset (Apache + F-Droid + GitHub). Values range between [0,1]. Numbers in bold represent the highest accuracy for each refactoring operation.}
\label{tab:rq1-results}
\end{table*}

\subsection{RQ1: \rqone}

In Table~\ref{tab:rq1-results}, we show the precision, recall, and accuracy of each ML algorithm in each one of the 20 refactoring operations, when training and testing in the 
entire dataset. Due to space constraints, 
we show the results of training and testing in individual datasets, as well as the confusion matrix, in our appendix~\cite{appendix}. 

\observation{Random Forest models are the most accurate in predicting software refactoring.}
Random Forest has the highest overall accuracy among all types of refactorings. Its average accuracy for class, method, and variable-level refactorings, when trained and tested in the entire dataset, are 0.93, 0.90, and 0.94, respectively. 
The only three refactorings that are below the 90\% threshold are \emph{Extract Class}, \emph{Extract and Move Method}, and \emph{Extract Method}.
Its average accuracy among all refactorings in all the datasets together, as well as Apache, GitHub, and F-Droid datasets only, are 0.93, 0.94, 0.92, and 0.90, respectively.
As a matter of comparison, the second best model is Decision Trees, which achieves an average accuracy of 0.89, 0.91, 0.88, and 0.86 in the same datasets.

\observation{Random Forest was outperformed only a few times by Neural Networks.}
In the F-Droid dataset, 
Neural Networks outperformed Random Forest 4 times 
(in terms of accuracy). 
Neural Networks also outperformed Random Forest in two opportunities in both the Apache and GitHub datasets.
However, we note that the difference was always marginal (around 1\%).

\observation{Naive Bayes models present high recall, but low precision.}
The Naive Bayes models presented recalls of 0.94, 0.93, 0.94, and 0.84 in the entire dataset, Apache, GitHub, and F-Droid datasets, respectively. These numbers are often slightly higher than the ones from Random Forest models, 
which were the best models (on average, 0.01 higher). 
Nevertheless, 
Naive Bayes models presented the worst precision values: 
0.62, 0.66, 0.62, and 0.67 in the same datasets. 
Interestingly, 
no other models presented such low precision.

\observation{Logistic Regression, as a baseline, shows good accuracy.}
Logistic Regression being, perhaps, 
the most straightforward model in our study, presents a somewhat
high overall accuracy, always outperforming Naive Bayes models. 
The average accuracy of the model in all the refactorings in the entire dataset is 0.83. 
Its best accuracy was in the \emph{Move Class} refactoring: 0.94 (which also presented high values in the individual datasets: in F-Droid, 0.94, in GitHub, 0.93, and in Apache, 0.95), and its worst accuracy, 0.77, was in the \emph{Extract and Move Method} and \emph{Inline Method} refactorings.
The overall averages are similar in the other datasets:
0.85 in Apache, 0.83 in GitHub, and 0.78 in F-Droid.

\subsection{RQ2: \rqtwo}

In Table~\ref{tab:rq2-results}, we show the most important features per refactoring level. The
complete ranking of features importance can be found in the online appendix~\cite{appendix}.

\observation{Process metrics are highly important in class-level refactorings.}
Metrics such as \textit{quantity of commits}, 
\textit{lines added in a commit}, and
\textit{number of previous refactorings} appear in the top-1 ranking very frequently.
In the top-5 ranking, seven out of the first ten features are process metrics; six out of the first ten are process
metrics in the top-10 ranking. Ownership metrics are also considered important by the models.
The \textit{author ownership} metric appears 32 times in the top-1 ranking; the
\textit{number of major authors} and \textit{number of authors} metrics also appear often
in the top-5 and top-10 rankings.

\observation{Class-level features play an important role in method-level and variable-level refactorings.}
Method-level refactoring models often consider class-level features 
(e.g., lines of code in a class, number of methods in a class) to be more important than method-level features.
In the top-1 ranking for the method-level refactoring models, 13 out of the 17 features are 
class-level features. In variable-level refactoring models, the same happens in 11 out of 17 features.
Interestingly, the most fine-grained feature we have, the \textit{number of times a variable is used} appears
six times in the top-1 ranking for the variable-level refactoring models.

\observation{Some features never appear in any of the rankings.}
For class-level refactoring models, the \textit{number of default fields}, and the \textit{number of synchronized fields}\footnote{By looking at the features distribution in our appendix~\cite{appendix}, we observe that most classes do not have synchronized fields; we discuss how feature selection might help in simplifying the final models in Section~\ref{sec:discussion}.} do not appear even
in the top-10 ranking. 
Nine other features never appear in the top-10 feature importance ranking of
method-level refactoring models (e.g., \textit{number of comparisons}, 
\textit{math operations}, and \textit{parenthesized expressions}), 
and ten features never make it in the variable-level refactoring models (e.g., 
\textit{number of loops}, and \textit{parenthesized expressions}).

\begin{table}
\small
\begin{tabular}{p{8cm}}
\toprule
\textbf{Class-level refactorings} \\
\textbf{Top-1:} quantity of commits (68), author ownership (32), lines added (6) \\
\textbf{Top-5:} quantity of commits (108), lines added (63), previous refactorings (63), author ownership (56), unique words in the class (47) \\
\textbf{Top-10:} quantity of commits (111), lines added (90), previous refactorings (90), unique words in the class (78), class LOC (70) \\
\midrule
\textbf{Method-level refactorings} \\
\textbf{Top-1:} class LOC (39), number of unique words in a class (15), number of methods in a class (13), class LCOM (9), number of fields in a class (6)\\
\textbf{Top-5:} class LOC (74), number of methods in a class (55), number of unique words in a class (52), class LCOM (37), number of final fields in a class (25) \\
\textbf{Top-10:} number of methods in a class (90), class LOC (88), class LCOM (71), number of unique words in a class (54), class CBO (54)\\
\midrule
\textbf{Variable-level refactorings} \\
\textbf{Top-1:} class LOC (27), class LCOM (10), number of unique words in a class (9), method LOC (7), number of public fields in a class (7)\\
\textbf{Top-5:} class LOC (61), number of unique words in a class (48), number of string literals in a class (38), number of variables in the method (30), number of public fields in a class (24)\\
\textbf{Top-10:} number of string literals in a class (72), class LOC (71), number of unique words in a class (66), number of variables in a class (55), number of variables in a method (49)\\
\bottomrule
\end{tabular}
\caption{Most important features for the models at different refactoring levels. Top-1, Top-5, and Top-10 indicate the number of times (in parenthesis) a specific feature appeared in the top-N ranking. For class and method level refactorings, a feature can at most appear 112 times; 96 times for a variable level refactoring. We show only the first five features per ranking; full list in the online appendix~\cite{appendix}.}
\label{tab:rq2-results}
\end{table}

\subsection{RQ3: \rqthree}

We show the precision and recall of each model and refactoring, 
in all the pairwise combinations of datasets in our
appendix~\cite{appendix}.
In Table~\ref{tab:rq3-results}, we show the overall average precision and recall of the Random Forest models
(the best model, according to RQ$_1$ results) when trained in one dataset and tested in another dataset.

\observation{Random Forest still presents excellent precision and recall when generalized, but smaller when compared to previous results.}
Random Forest models achieve precision and recall of 0.87 and 0.84, when trained using the GitHub repository, 
the largest repository in terms of data points, 
and tested in Apache. 
When trained in the
smallest dataset, F-Droid, Random Forest still performs reasonably well: precision and recall of 0.77 and 0.73 when tested in Apache, and 0.81 and 0.76 when tested in GitHub. 
Nevertheless, we remind the reader that in terms of accuracy,
Random Forest achieved average scores of around 90\%. 
In other words, models seem to perform best when trained with
data collected from different datasets.

\observation{Method and variable-level refactoring models perform worse than class-level refactoring.}
In general, class-level refactoring models present higher precision and recall than the method- and
variable-level refactoring models. 
Using a model trained with the GitHub data set and tested in the F-Droid
data set, the average precision and recall for Random Forest models at class-level are 0.92 and 0.92.
On the other hand, the average precision and recall for Random Forest models at method-level are 0.77 and 0.72, 
respectively; at variable-level, we observe precision and recall of 0.81 and 0.75.

\observation{SVM outperforms Decision Trees when generalized.}
We observed Decision Trees being the second best model
in RQ$_1$.
When carrying models to different contexts, however, we observe that SVM is now the second best model, and
only slightly worse than the Random Forest. For example, in the appendix, we see that for a model trained in GitHub and tested in Apache, the average precision and recall of SVM models is 0.84 and 0.83 (in contrast, Random Forest models have 0.87 and 0.84). The difference between both models is, on average, 0.02.

\observation{Logistic Regression is still a somewhat good baseline.}
Logistic Regression baseline models, when carried to different contexts, still present somewhat good numbers.
As an example, the models trained with GitHub data and tested in the Apache dataset show an average precision and recall of 0.84 and 0.83. The worst averages happen in the models trained with the Apache dataset and tested in
the F-Droid dataset (precision of 0.75 and recall of 0.72).

\observation{Heterogeneous datasets might generalize better.}
More homogeneous datasets (i.e., the Apache and F-Droid datasets), when carried to other contexts, present lower precision and recall.
This phenomenon can be seen whenever Apache and F-Droid models are cross tested; their precision and recall never
went beyond 0.78. This phenomenon does not happen when GitHub, a more heterogeneous dataset in terms of different domains and architectural decisions, is tested on the other two datasets. 

\begin{table}
\small
\centering
\begin{tabular}{lrr|rr|rr}
\toprule
 & \multicolumn{2}{c|}{\textbf{Apache}} & \multicolumn{2}{c|}{\textbf{GitHub}} & \multicolumn{2}{c}{\textbf{F-Droid}} \\
 & \textbf{Pr}     & \textbf{Re}     & \textbf{Pr}     & \textbf{Re}     & \textbf{Pr}      & \textbf{Re}     \\
\midrule
\textbf{Apache}        & - & -      & 0.84          & 0.79       & 0.77           & 0.70       \\
\textbf{GitHub}        & 0.87          & 0.84       & - &-       & 0.84           & 0.80       \\
\textbf{F-Droid}       & 0.77          & 0.73       & 0.81          & 0.76       & - & -      \\
\bottomrule
\end{tabular}
\caption{The average precision (Pr) and recall (Re) of the 20 refactoring prediction Random Forest models, when trained in one dataset and tested in another dataset. Rows represent datasets used for training, and columns represent datasets used for testing.}
\label{tab:rq3-results}
\vspace{-8mm}
\end{table}

\section{Discussion}
\label{sec:discussion}

In the following, 
we extensively discuss some important ramifications of our research.
More specifically, we discuss:

\begin{enumerate}
\item the challenges in defining $k$ as a constant to collect non-refactored instances,
\item the features used for model building as well as their interpretability,
\item the importance of process and ownership metrics (and the need for fine-grained metrics),
\item the need for larger and more heterogeneous datasets to achieve higher generalizability,
\item how to prioritise the refactoring recommendation suggestions given by the models,
\item the need for more fine-grained refactoring recommendations,
\item the recommendation of high-level refactorings,
\item taking the developers' motivations into account,
\item the use of Deep Learning (and Natural Language Processing algorithms) for software refactoring, and
\item the challenges of deploying ML-based refactoring recommendation models in the wild.
\end{enumerate}

\subsection{Collecting non-refactored instances via an heuristic} 
\label{sec:non-refactored-instances}

The identification of negative instances, i.e., code elements that did not undergo a refactoring operation, is an important theoretical problem that our research community should overcome.

We propose the use of code elements that did not undergo refactoring operations for $k$ commits in a row. In this particular paper,
we chose $k=50$ (i.e., 50 commits in a row without being refactored)
as a constant to determine whether a class, its methods, and its variables should
be considered an instance of a non-refactoring. The number 50 was chosen 
after manual exploration in the dataset.

To measure the influence of $k$ in our study, we re-executed our data collection procedure
in the entire dataset (11,149 projects) with two different values for $k$:

\begin{itemize}
    \item $k=25$. The half of the value used in the main experiment. A threshold of 25 means
    that we are less conservative when considering instances for the non-refactoring dataset.
    In this dataset, we have a total of 7,210,452 instances (at class, method, and variable levels).
    This represents an increase of 7.1 times when compared to the dataset in the main experiment.
    
    \item $k=100$. The double of the value used in the main experiment. A threshold of 100 means
    that we are more conservative when it comes to considering a class as an instance of a 
    non-refactoring. In this dataset, we have a total of only 120,775 instances. This represents
    around 12\% of the dataset in the main experiment.
\end{itemize}

We note that the distribution of the features values of the non-refactored instances 
in $k=25$ and $k=100$ datasets are somewhat
different from each other. As examples, the quantiles of the \emph{CBO at class-level} in $k=25$ dataset are [1Q=17, median=35, mean=57, 3Q=69], whereas the quantiles in $k=100$ dataset are [1Q=6, median=28, mean=54, 3Q=75];
for the \emph{WMC at class-level}, we observe, [1Q=59, median=145, mean=273, 3Q=343]for $k=25$, and [1Q=72, median=266, mean=425, 3Q=616] for $k=100$; for the \emph{LOC at class-level}, we observe [1Q=320, median=734, mean=1287, 3Q=1626] for $k=25$, and [1Q=466, median=1283, mean=1568, 3Q=2189] for $k=100$.

We trained Random Forest models (given that it was the algorithm with the best accuracy in RQ$_1$) in both $k=25$ and $k=100$ datasets.
In $k=25$, the average of the absolute difference in the precision and recall of the 20 refactoring models,
when compared to $k=50$, are 0.0725 and
0.099, respectively. In $k=100$, the average of the absolute difference in precision and recall when compared
to $k=50$ are 0.0765 and 0.064, respectively. The precision and recall of each refactoring is in our online appendix~\cite{appendix}.

In $k=25$, however, in only four (\emph{Move Class}, \emph{Move and Rename Class}, \emph{Extract Method}, 
and \emph{Rename Method}; out of 20) models, the precision values were better
than in the $k=50$. Similarly, only a single model (\emph{Rename Method}) 
had a better recall when compared to $k=50$.
This might indicate that $k=25$ is not a good threshold, as it might be too small.
In $k=100$, while it is hard to distinguish whether it has a better precision than
$k=50$ (11 models did better with $k=50$ and 9 models did better with $k=100$),
models in $k=100$ had almost always a better recall (16 models out of 20).
This might indicate that more conservative thresholds might help
in increasing recall at the expense of precision performance.
This discussion shows the importance of finding the right threshold to determine classes, methods, and variables that can serve as non-refactoring instances.

It is worth emphasizing that our proposed heuristic to detect counterexamples
of refactoring instances is an approximation. 
While we believe that our assumption that classes that can still be evolved by developers without the need for refactoring can serve as good counterexamples for the model, these data points are simply approximations. 
There might be other more effective counterexamples that would contribute to the creation of better models.

As an alternative, when designing the model, we considered the possibility of doing the extraction of non-refactored instances at commit-level. For example, 
whenever a refactoring $R$ was detected in a class, method, or variable $a$, 
we extracted all the other elements that existed in the modified files of that commit as examples of non-refactored instances. 
We relied on the assumption that the elements that did not change in that commit could be used as counterexamples during model creation.
We, however, discarded this idea after some exploration. 
When looking at individual commits \textit{only} and not at larger time windows, 
one can not determine whether a code element is an example of an element that does not need to be refactored. 
The same code element might have changed in the subsequent commit, 
thus rendering the previously collected data invalid (i.e., mischaracterizing the counterexample).
Furthermore, 
another factor we took into account was that, if we consider all elements that were not refactored in every single commit as a counterexample, 
the number of extracted data points would be orders of magnitude higher than the number of data points for refactoring instances. 
That would make the dataset highly imbalanced.
We decided not to deal with a highly imbalanced datased because that is a known challenge in ML~\cite{he2009learning}.

Given that the current state-of-the-art enables us to precisely identify refactoring operations that have happened in software systems, but the identification of non-refactoring instances is challenging, 
we suggest the possibility of training models using solely a single class. 
In this case, one would train the model solely on the real-world refactoring instances, and use the outcome probability of the model to decide whether to recommend a refactoring operation. 
We expect models to return a very low probability in methods that do not need to undergo refactoring.
Note that, in this way, there is no need for collecting non-refactoring data points, 
which would avoid the problem discussed in the previous subsection. 
We therefore suggest future work to explore the performance (as well as the drawbacks) of such models in recommending software refactoring.

Nevertheless, 
the fact that our community does not have an accurate dataset composed of examples of code elements that do not need refactoring is a threat to any study in software refactoring. 
Our community has been working on several approaches to point developers to problematic pieces of code for a long time. 
However, 
less research has been dedicated to revealing exemplary pieces of code (exemplary in the sense that 
these pieces of code do not warrant refactoring operations). 
Given the data-driven era we find ourselves in, 
research investigating the identification and creation of a sample of such pieces of code might be highly relevant.

\subsection{Features and their interpretability}

Our models use a set of source code, process, and ownership metrics as features (see
Table~\ref{tab:features} for the complete list). The choice of features was mostly based on previous ML models for software engineering tasks (e.g.,~\cite{scalabrino2017automatically,mcintosh2014impact,d2012evaluating}).

Our conjecture when we settled on using structural metrics was that the structure of a class or method is an important factor that developers take into consideration when identifying pieces of code to refactor, e.g., a complex method is much more prone to being refactored than a structurally less complex method. 
Given the high accuracy, precision, and recall that we observed in our empirical study, our conjecture seems to hold. 
We understand that some of the metrics might seem counterintuitive. 
Some developer might be hard-pressed to explain why something as the \emph{number of mathematical operations} in a given part of the code may indicate that refactoring is warranted.

Given the amount of features that are readily available and that have been used in the literature, 
we decided not to perform manual feature selection (i.e., manually selecting the most appropriate features given the data and the model). 
Rather, we decided to let the model decide which ones have more predictive power.

Interestingly, as the results we used to answer our RQ$_2$ seem to suggest, 
models tend to selected features that also make more sense to humans. 
For example, number of methods in a class (which was chosen 13 times as the most important predictor in method-level refactorings) or number of lines of code in a class (which was selected 39 times as the features that most contributed to model building). 
On the other hand, the number of parenthesized expressions and number of lambdas do not seem to help models in learning how to recommend refactoring; such features were automatically discarded (i.e., never used) by these models.

We, nevertheless, understand that the interpretability of these models can play a decisive factor in whether developers will accept the recommendation. 
Developers might want to know why the model is suggesting a specific refactoring. 
While interpretability of models is a complex problem in the area of ML in general~\cite{rapp2019simplifying,kim2017interpretability}, making use of metrics that developers can better relate to, 
as well as showing them what metrics most influenced the model to recommend a given refactoring might make the developer more confident in accepting the recommendations. 
Interpretability of refactoring recommendation models is therefore an important future work. 

\subsection{The importance of process and ownership metrics (and the need for fine-grained metrics)}
\label{subsec:importance-of-other-metrics}

We observed that process metrics are indeed considered important by the models (see RQ$_2$). 
For example, the \emph{number of commits} metric figured as the most important feature in the class-level refactoring models. 
Additionally, 
related research suggests that defect prediction models~\cite{rahman2013and,madeyski2015process,d2012evaluating} also benefit from process metrics. 

While source code metrics are able to capture the structure of a code element, 
process metrics are able to capture its evolution history (e.g., number of changes, code churn, number of bugs, or refactorings over time). Such characteristics seem to play an important factor when deciding whether to refactor the code element. We would argue that this is inline with general knowledge on software design. For example, changing a class several times eventually leads to brittle design (following Lehman's laws of software evolution~\cite{Lehman1997,lehman1996laws}), which drives developers to remedy the situation by refactoring the class; or a class that has presented a high number of bugs tend to require more ``clean up'' than classes that do not suffer from the same issue. Process metrics, thus, provide models with a perspective on the evolution of the class.

We currently use process and ownership metrics to support the prediction of class-level refactorings only. These metrics are naturally collected at file-level, and collecting them in a more fine-grained manner (i.e., method and variable ownership) would require complex tooling to be developed. 
We can only conjecture that process metrics would also help models in better predicting refactoring at method and variable levels. To that aim, it is our goal to 
\begin{enumerate*}[label=(\roman*), font=\itshape]
\item develop a tool that is able to collect process and ownership metrics at method and variable levels,
\item feed our models with these new features,
\item re-execute our ML pipeline and examine how accuracy is affected.
\end{enumerate*}

In addition, D'Ambros et al.~\cite{d2012evaluating} observed that the number of bugs, when extracted by means of string matching in the commit message (which is our case), might reduce the quality of the resulting predictor. 
Our models currently indicate that the ``number of bugs'' feature is relevant. 
This feature frequently appeared in the top five and top ten ranking of features that most contributed to model building. 
We surmise that a more precise approach to detecting and counting bugs, 
which might require better integration with issue tracking systems, will improve the quality of the recommendations.

\subsection{Making a case for larger and more heterogeneous datasets}

According to the results we used to answer RQ$_3$, 
larger and more heterogeneous datasets tend to generalize better. 
We would argue that large amounts of diverse refactoring operations contribute to the creation of stronger, more accurate models. 

While our dataset might be already considered a large one, 
with around 3 million labelled instances, we believe that the collection of even larger datasets compressed of different types of systems and refactoring operations will 
result in more helpful models able to provide developers with more accurate refactoring recommendations. 
Moreover, 
it is a common observation in ML studies (not only in software engineering tasks) that simple models trained on large datasets often work better than complex models trained on small datasets~\cite{code2vec, mikolov2013efficient}. 
Simple models are cheaper to train and store.

\subsection{Prioritizing refactoring recommendations}

All our models currently perform binary classification. 
In other words, each model is only able to predict a single refactoring operation. 
Our empirical study shows that, when tested in isolation, models have high accuracy. 

We envision a recommendation tool making use of all the models together in order to recommend all the possible refactorings. 
Suppose we want to offer refactoring recommendations for a given method, we would need to pass the method through the seven different method-level refactoring models; each of these seven models would give its own prediction, 
and we would show the resulting list of recommendations to the developer.

We understand that in a scenario in which developers are faced with lots of refactoring recommendations it might be hard for them to work out which refactorings to prioritize. 
An avenue to explore in future work is to take advantage of the probability values that are internally produced by 
the models to prioritize which refactorings are more appropriate in a given context. 
A tool that presents these probability values to the end-users could allow them to decide which refactorings 
they should apply and in which order 
(we discuss more usability concerns of such a tool in Section~\ref{subsec:models-in-the-wild}).

\subsection{The need for more fine-grained refactoring recommendations}

In this first step, 
we have showed that ML can model 
the refactoring recommendation problem. 
Although the current models provide recommendations
at different levels of granularity (i.e., class, method, and variable levels), 
there is room for improvement by fine-tuning models to offer even more fine-grained refactoring recommendations.
Take as an example
the \textit{Extract Method} refactoring. Our models can identify which method
would benefit from an extraction; however, it currently does not point to
which parts of that method should be extracted (i.e., initial token and end token). Another example are refactorings that involve more than one class, e.g., \textit{Move Method} or \textit{Pull Up Method}: to which class should
the method be moved to?

We see a future where, for each of the refactorings we studied, a highly-specific model, able to provide fine-grained recommendations, is devised. We conjecture that
models that learn precisely, e.g., what tokens to extract out of a method,
would need to be deep. Therefore, we believe that deep learning will play an important role in the field of software refactoring in the near future.
We discuss deep learning later in this section.

\subsection{The recommendation of high-level refactorings}

In this study, we explore recommendations of low-level refactorings, i.e., small and localized changes that improve the overall quality of the code. We did not explore recommendations
of high-level refactorings, i.e., larger changes
that improve the overall quality of the design. 

We see that the great challenge of recommending high-level refactorings is that
the model requires even more context to learn from. Before applying a design pattern to the source code, developers often think about how to abstract the problem in such a way that the pattern would fit. 

As an initial step, the book of Kerievsky~\cite{kerievsky2005refactoring} might
serve as a guide.
In his book, the author shows how to move code, that is often implemented in a procedural way, to a
design pattern oriented solution, by means of low-level refactorings. Our next step
is to explore how we can ``aggregate'' several low-level refactoring recommendations
in order to provide developers with high-level refactoring suggestions.

\subsection{Taking the developers' motivations into account}

Empirical research shows that developers refactor for several reasons, other than to ``only improve the quality of the code'' (e.g.,~\cite{kim2012field,silva2016we}). 
In our first foray into applying ML algorithms to predict refactoring operations, 
our models do not factor in ``motivation''. 
Nevertheless, note that our large dataset of refactorings contains refactorings that have happened for varying reasons (given that we never filtered refactoring based on motivation from the projects). 
Interestingly, 
our models still show high accuracy. 
Exploring whether models built specifically for, e.g., ``refactoring to add new functionality'', 
would provide even better results, not only in terms of accuracy, but also in terms of ``developer satisfaction''.
We defer this development to future work.

\subsection{The use of Deep Learning (and Natural Language Processing algorithms) for software refactoring}
\label{sec:use-of-nlp} 
Programming languages have phenomena like syntax and semantics~\cite{allamanis2018survey,Hindle:2016:NS:2930840.2902362, Hellendoorn:2017:DNN:3106237.3106290}. 
Motivated by several recent works that use advanced ML algorithms on source code with the goal of (semi) automating several non-trivial software engineering tasks such as suggesting method and class names \cite{Allamanis_2015}, code comments \cite{Wong_2013}, generation of commit messages \cite{jiang2017automatically}, and defect prediction \cite{Deep_2017},
we intend to experiment NLP-specific deep learning architectures to deal with code refactorings. Using models like Seq2Seq~\cite{Sutskever2014SequenceTS} and
Code2Vec~\cite{code2vec}, both refactorings predictions and refactored code can be outputs of the model, having the source code only as input. 
To facilitate the work of future researchers interested on the topic, our online appendix~\cite{appendix} contains a dataset with all the refactored classes studied here.

\subsection{Refactoring recommendation models in the wild}
\label{subsec:models-in-the-wild}

As mentioned, 
popular tools such as PMD and Sonarqube
offer detection strategies for common code smells, e.g., \emph{God Classes} and \emph{Long Methods}.
These tools have been integrated into different stages of the developers' workflow, e.g.,
inside IDEs, during code review, or their results have been incorporated into quality reports.
We envision a ML-powered refactoring recommendation tool finding its way into the daily life of a developer in the same way linters and code quality recommendation tools (e.g., PMD, Checkstyle, Sonarqube) currently belong to their daily routine. 
However, the deployment of ML-based refactoring recommendation models does not come without
its challenges. 

First, prediction models take up a lot of disk space (some of the models we built throughout this research take up around 700MB to 1GB of disk space), 
which makes them unwieldy to deploy inside IDEs (without mentioning that loading them into memory would require sizable memory resources). 
While the ML research field is still looking for efficient ways of compressing large models (see~\cite{bucilua2006model} for details), 
introducing the 20 ML models that we built into the developers' machines/IDEs  is certainly not a feasible solution. 
A possible workaround to this challenge would be to provide a centralized server that provides recommendations to clients (e.g., IDEs and code review tools). 

Another way to reduce the size of our models would be to build leaner models. 
In RQ$_2$, we show that some features never make to the top-10 ranking features; others were never even used. 
Future work should investigate which features can be removed without significant loss of prediction power, 
thus on removing features that have no real prediction power, and on identifying the simplest model that works by, e.g., performing feature reduction. As a reference, we refer the reader to Kondo et al's work~\cite{kondo2019impact}. Authors explored the impact of eight different feature reduction techniques on defect prediction models; we suggest the same line of work for refactoring recommendation models.

Moreover, 
our empirical study shows that the training of these models take hours 
(some of our Random Forest models took approximately 2 hours running on a machine equipped with a 40-core processor).
On the other, once these models are trained, prediction happens almost instantly.
This is due to the fact that our models require a feature vector composed solely 
by code metrics that are easily extracted from source code. 
The long training time
reinforces the need for generalizable models (which are possible to obtain according to our results), 
given that many companies are not able to afford the costly model training.

Second, program analysis tools solely require access to the source code of
the program for the recommendation to happen. Our current ML models 
also require ownership and process metrics. 
While our results show that these
metrics play an important role in the models, they are less trivial to be calculated, 
requiring access to the full history of the project as well as maintenance.
Future work should evaluate what to do in cases where the developer does not
have access to these metrics, i.e., when offering consultancy to a company that does not
provide the consultant with the full repository, 
or when the project is in earlier stages 
and the repository still does not contain useful data. 

Third, the usability aspects that such a tool would need in order for developers
to trust it.
In this paper, we do not explore such aspects.
While this is not unique to ML-based recommendation models, we believe this is
an important aspect to be explored, given that the interpretability of these (black box) models
are harder than the detection strategies our community currently relies on.
Guidelines on how to recommend software refactoring~\cite{bavota2014recommending} as well
as lessons learned on building large-scale
recommendation tools~\cite{sadowski2015tricorder,sadowski2018lessons} are of
great help. 
Given that ML models are drawing a lot of attention from the software engineering community,
other researchers have already started to probe into the usability-related issues of ML-powered solutions 
to software engineering tasks (e.g., \cite{kumar2019building}).

Finally, understanding 
whether it is possible for a company to reuse existing models 
(a practice commonly used in other communities, 
such as the reuse of pre-trained models as Word2Vec~\cite{goldberg2014word2vec} and BERT~\cite{devlin2018bert})
and
how often the refactoring recommendation models should be re-trained are fundamental questions that still need to be answered.

\section{Related Work}\label{sec:relatedWork}

After the publication of Fowler's seminal book \cite{Fowler1999-go}, 
refactoring went mainstream and many surveys and literature reviews on the 
subject were performed. 
One of the early surveys that brought refactoring into the limelight of researchers was carried out by \citet{a_survey_of_software_refactoring}: 
their survey is centered around refactoring activities, supporting techniques, and tool support.
Specifically, 
their discussion is organized around software artifacts and how refactoring applies to them, 
so the authors emphasize 
requirement refactoring, 
design refactoring, and 
code refactoring.
Additionally, \citeauthor{a_survey_of_software_refactoring} briefly share
their outlook on the impact of refactoring on software quality. 
Their survey, however, 
took only a few studies on identifying refactoring opportunities into account and 
did not follow a systematic approach. 
As mentioned, 
since then several systematic literature reviews have been conducted on refactoring. 

The existing literature discusses different automatic refactoring approaches whose purpose is helping practitioners 
in detecting code smells, 
some of which are even able to suggest the refactoring activities that should be performed by the practitioners in order to 
remove the detected code smells. 
Most approaches are either based on rules, 
employ search-based algorithms, 
or ML approaches. 
A recent systematic literature review \cite{automatic_software_refactoring_a_slr} shows that there has been an increase in the 
number of studies on automatic refactoring approaches. 
According to the results of such literature review, source code approaches have been receiving more attention from researchers than model based approaches.
In addition, the results indicate that search-based approaches are gaining increasing popularity and 
researchers have recently begun exploring how ML can be used to help practitioners in 
identifying refactoring opportunities.
The concepts and rule-based approaches proposed by early researchers that laid the theoretical foundation for more recent advances in the area 
are presented in Subsection~\ref{subsec:code_smell_detection}. 
Related work on search-based approaches applied to refactoring is discussed in Subsection~\ref{subsec:search_based_refactoring} and 
related work on ML is reviewed in Subsection~\ref{subsec:machine_learning}.

\subsection{Code smell detection}\label{subsec:code_smell_detection}

In hopes of providing a insightful understanding of 
code smells, 
the goals of studies on code smells, approaches used to probe into code smells, 
and evidence that bolsters the fundamental premise that code smells are symptoms of issues in the code, 
\citet{code_bad_smells_a_review_of_current_knowledge} carried out a systematic literature review in which they 
synthesized the results of 39 studies on code smells. 
Since we consider the identification of code smells and the detection of 
refactoring opportunities two related problems, 
it is also worth mentioning the systematic literature review performed by 
\citet{identifying_refactoring_opportunities_in_oo_code_slr}. 
\citeauthor{identifying_refactoring_opportunities_in_oo_code_slr} discusses studies that consider both code smells and refactoring opportunities from a different perspective: 
the main focus of their literature review is providing an overview of code smell identification approaches. 
Based on an analysis of 47 studies, 
\citeauthor{identifying_refactoring_opportunities_in_oo_code_slr} concluded that although 
there was a sharp increase in the number of studies on identifying refactoring opportunities, 
up to 2013 the results of these studies were derived mostly from relatively small datasets.
\citet{refactoring_for_disclosing_code_smells_in_oo_software} extended the systematic literature review carried out by \citeauthor{identifying_refactoring_opportunities_in_oo_code_slr} focusing on code smells identification and anti-patterns. 
The two main contributions of their survey is highlighting the datasets and the tools employed in the selected studies and 
the identification of the code smells that were most used in these studies. 

Recently, 
\citet{a_systematic_review_on_the_code_smell_effect} performed a systematic literature review to summarize knowledge about how 
code smells impact software development practices, 
which the authors termed ``smell effect''. 
\citeauthor{a_systematic_review_on_the_code_smell_effect} selected and analyzed 
64 studies that were published between 2000 and 2017.
One of the main findings reported by the authors is that 
human-based evaluation of smells is not reliable: 
a trend in the selected studies seems to indicate that developers have a low level of consensus on smell detection. 
Furthermore, 
their analysis of the selected studies suggests that demographic data as developers' experience can significantly impact code smell evaluation. 

\citet{a_review_based_comparative_study_of_bad_smell_detection_tools} carried out a systematic literature review on code smell detection tools. 
Their study is centered around the identification of code smell detection tools, 
their main features, and the types of code smells that these tools are able to identify. 
\citeauthor{a_review_based_comparative_study_of_bad_smell_detection_tools} also performed a comparison of the four most widely used tools (i.e., most frequently mentioned in the selected studies). 
It is worth mentioning that considering the selected studies, which were published from 2000 to 2016, 
no tool implements a ML based approach: 
this indicates that only recently researchers have begun investigating ML models in this context. 
\citet{a_review_of_code_smell_mining_techniques} also performed a systematic literature review on tools and approaches to 
mining code smells from the source code. 
Essentially, 
\citeauthor{a_review_of_code_smell_mining_techniques} classified tools and approaches based on their detection methods. 
\citeauthor{a_review_of_code_smell_mining_techniques} emphasized mining approaches, thus they did not take ML-based approaches into account. 

\subsection{Search-based refactoring}\label{subsec:search_based_refactoring}

\citet{mariani2017systematic} carried out a systematic literature review of how search-based approaches have been applied to refactoring. 
\citeauthor{mariani2017systematic}
found that evolutionary algorithms and, in particular, genetic algorithms were the most commonly used algorithms in the analyzed studies.
In addition, 
they found that the most widely used and investigated refactorings are the ones in Fowler's catalog~\cite{Fowler1999-go}.
More recently, 
\citet{a_survey_of_search_based_refactoring_for_software_maintenance} also looked at search-based refactoring. 
However, 
differently from the literature survey by \citeauthor{mariani2017systematic}, 
\citeauthor{a_survey_of_search_based_refactoring_for_software_maintenance} give a more in-depth review of the selected studies
in the sense that \citeauthor{a_survey_of_search_based_refactoring_for_software_maintenance} also cover other aspects of the literature. 
For instance, 
\citeauthor{a_survey_of_search_based_refactoring_for_software_maintenance} also discuss the tools used in the selected studies as well as 
provide an investigation of how some metrics have been tested and discussed in the selected literature. 
In addition, 
\citeauthor{a_survey_of_search_based_refactoring_for_software_maintenance} detail how the search-based approaches described in the selected studies have evolved over time. Similarly to the results presented by \citeauthor{mariani2017systematic}, 
\citeauthor{a_survey_of_search_based_refactoring_for_software_maintenance} also found that evolutionary algorithms are the most commonly used algorithms in the selected studies.

\subsection{ML algorithms}\label{subsec:machine_learning}

To our best knowledge, 
only one systematic literature review~\cite{machine_learning_techniques_for_code_smell_detection} has been conducted with the purpose of summarizing the research on ML 
algorithms for code smell prediction.
\citeauthor{machine_learning_techniques_for_code_smell_detection} selected 15 studies that describe code smell prediction models. 
\citeauthor{machine_learning_techniques_for_code_smell_detection} analyzed the selected studies in terms of \emph{(i)} code smells taken into account, 
\emph{(ii)} setup of the ML based approaches, 
\emph{(iii)} how these approaches were evaluated, and 
\emph{(iv)} a meta-analysis on the performance of the code smell prediction models described in the selected studies. 
According to the results, 
God Classes, Long Methods, Functional Decomposition, and Spaghetti Code are the most commonly considered code smells. 
Decision Trees and SVM are the most widely used ML algorithms for code smell detection. 
Additionally, 
JRip and Random Forest seem to be the most effective algorithms in terms of performance.

\section{Threats to validity}
\label{sec:ttv}

This section outlines the threats to the validity of our study. 

\subsection{Construct validity}
Threats to construct validity concern the relation between the theory and the observation, and in this work are mainly due to the measurements we performed.

\begin{itemize}

\item Our strategy for gathering the large amount of data we investigated entailed 
mining a large number of software repositories for instances of class, method, and variable refactorings.
Thus, 
the main internal validity threat is the data collection process. 
We cannot rule out the issues that arise when performing large scale data extraction (issues indeed
happened, as discussed in Section~\ref{sec:execution}). 
We provide a replication package containing all experimental scripts and datasets used in our study so that 
researchers and practitioners can fully replicate and confirm our results. 

\item As mentioned in Section~\ref{sub:feature_extraction}, 
over the course of our data extraction process, we determined the number of bug fixes by employing a keyword matching approach. 
The approach is widely used by the mining software repositories community to detecting bug fix related information in software repositories. 
It is worth noting that the effectiveness of such approach depends on the keywords used during the data extraction process, 
so we acknowledge the possibility that we might have overlooked the inclusion of relevant keywords.

\item
Our data collection mechanism makes use of RefactoringMiner~\cite{Tsantalis_2018}, 
a tool that is able to identify refactoring operations in the history of a repository. 
Therefore, the soundness of our approach hinges on the effectiveness of refactoring detection tool we used. 
RefactoringMiner presents a precision and recall of 98\% and 87\%, respectively, in detecting the refactoring operations we study.
We did not re-evaluate the precision and recall of RefactoringMiner in the studied sample, as this was already established in their research.
Given how RefactoringMiner works internally and that RefactoringMiner was evaluated on projects with similar characteristics (in fact, 65\% of the projects in RefactoringMiner's evaluation dataset are in our dataset), we have no reason to believe that the accuracy reported in the literature would not apply to our study.

\item
An underlying assumption of our research is that refactorings that have happened in the past are good examples of refactorings that will happen in the future. 
Our models never learn from ``refactorings that developers find to be important, but never got around to carrying them out''.
Nevertheless, 
given the amount of data points we use for training, 
we have no reason to believe that ``refactorings that developers consider relevant but ended up never being carried out'' are so intrinsically different from the ``refactorings that developers carried out''. 
In ML terms, 
we do not believe their feature vectors would have such a different distribution that models would not be able to predict them with a reasonably good accuracy. 
This is, however, a conjecture.
Case studies in industrial settings, 
in which developers annotate not only whether the recommendations of ML-based models were pertinent, 
but also refactorings they would like to perform in elements that our models do not identify refactoring opportunities, 
is a necessary step in order to test this conjecture.

\item
Finally, one of the metrics we also used to train our models was the ``number of default methods'' (at class-level). However, later in one of our inspections, we observed that, due to a bug in the metric collections tool, the number of default methods was always zero.\footnote{\url{https://github.com/mauricioaniche/ck/commit/f60590677271fb413ecfb4c2c5d0ffbaf8444075}.} All the learning algorithms ignored this metric, as it indeed added no value to the learning process; in fact, it appeared on the list of features that were never used by our model. Therefore, we affirm that this bug does not influence the overall results of our paper. Moreover, we have no reason to believe that the adding this feature would bring significant improvements. Nevertheless, we propose researchers to use this feature in future replications of this paper. 
\end{itemize}

\subsection{Internal validity}
\label{sec:internal-validity}

Threats to internal validity concern external factors we did not consider that could affect the variables and the relations being investigated. 

\begin{itemize}

\item 
We removed projects that failed during data collection. As we discussed in Section~\ref{sec:sample}, our pipeline is composed of several tools, all of them being prone to failures, e.g., RefactoringMiner running out of memory. The percentage of failed projects is small (8\%), and does not affect the representativeness of our final dataset.

\item
Owing to the fact that, in most cases, 
there are more instances of the non-refactoring class in our dataset 
than instances of the refactoring class (see Table~\ref{tab:sample-refactorings}), 
we had to cope with an imbalanced dataset. 
Given that there is no reliable estimate on the distribution of refactoring and non-refactoring instances ``in the wild'', 
we decided to perform under-sampling.
That is, we chose to remove instances from the over-represented class by means of random under-sampling. This means that the
dataset for each refactoring operation is bounded by the minimum between the number of positive and the number of negative instances
(e.g., as we see in Table~\ref{tab:sample-refactorings}, although we have 41,191 instances of \textit{Extract Class} refactorings,
we have only 10,692 instances of non-refactored classes, and thus, our model is trained on all the 10,692 instances of 
non-refactored classes + 10,692 randomly sampled instances of refactored classes).

To better measure the impact of this choice, we re-created the Random Forest models using a ``Near Miss''
under-sampling strategy~\cite{mani2003knn}. While the average absolute difference in precision and recall
are 0.116 and 0.052, respectively, it is hard to distinguish which strategy helps the model in improving
accuracy. In nine out of 20, Near Miss improved the precision when compared to the random sampling strategy (and thus, random performed better in 11 models), whereas in 12 out of 20, Near Miss improved the recall. As 
we conjecture that, in the refactoring problem, classes will always be unbalanced by nature, future research
is necessary to better understand how to under (or even over) sample.
Nevertheless, we acknowledge that a balanced dataset may be different from the distribution that is expected in real life. 
Hence, 
a balanced dataset has the potential to lead to less accurate models in practice. 
Exploring the performance of models trained on datasets that reflect reality is important future work that should be tackled once, as a community, we understand what the real distribution is.

\item
Our ML pipeline performs scaling and undersampling. Improving the pipeline, e.g., by applying better feature reduction, different balancing strategies, and extensive hyperparameter search, will only make our results better.
While developing production-ready models was not the main goal of this paper,
we note that our open-source implementation available in our appendix~\cite{appendix} enables it effortlessly. In other words, any researcher or company can download our implementation and datasets, use their available infrastructure, and train (even more accurate) models.

\item
Code smells are symptoms that might indicate deeper problems
in the source code~\cite{Fowler1999-go}. 
While code smells have been shown to greatly indicate problematic pieces
of code, in this work, we did not use them as features to our model. However, we note that code smells are detected by combinations of proxy metrics (i.e., detection strategies~\cite{marinescu2004detection}). These proxy metrics are commonly related to the structure of the source code (e.g., complexity/WMC, coupling/CBO) and are highly similar to the structural metrics we use as features (see Table~\ref{tab:features}). In other words, we train our models with metrics that are similar to the metrics used by the code smells detection strategies. Therefore, we conjecture that using code smells as features would add only a small amount of information for the models to learn from. That being said, making sure that all catalogued code smells are covered by our features is interesting future work which might increase the accuracy of refactoring recommendation models. Our own previous research shows that code smells might be architecture-specific~\cite{aniche-1,aniche-2,aniche-3,aniche-4}, e.g., MVC systems might suffer from different and specific smells than Android systems.

\item
As we discuss in Section~\ref{sec:discussion}, we did not take into account
the different reasons a developer might have when deciding to refactor, e.g.,
to add a new functionality, or to improve testability. These motivations might
indeed change (or even help the developer to prioritize) which refactors
to apply. Nevertheless, we affirm that the goal of this first study was to explore
whether ML can model the refactoring recommendation problem. Given that we observed
high accuracy, we can only conjecture that taking the motivation into consideration will only increase the accuracy (or again, help in prioritization) of the models. We leave it as future work.

\item
We consider our dataset as a set of unordered refactorings. As a contrast, studies in defect prediction consider datasets as a set of ordered events, e.g., they do not mix ``past'' and ``future'' when evaluating the accuracy of their models (e.g., \cite{pascarella2019fine}). We argue that there is no need for such design, given that we devise a single cross-project model, based on hundreds of thousands of data points from more than 11k projects altogether. In other words, we do not devise one model per project, as commonly done in defect prediction. Thus, we affirm that the model has little chance of memorizing specific classes. Our 10-fold random cross validation (and the individual precision and recall of each fold, that can be seen in our appendix~\cite{appendix}) also gives us certainty that this is not a threat. Nevertheless, to empirically show that our decision of not ordering the dataset does not influence our results, we trained Random Forest models using the first 90\% of refactorings that have happened (ordered by time) and tested on the remaining 10\% of refactorings that
happened afterwards. We obtained an average accuracy of 87\% among the 20 refactorings. The individual results per type of refactoring can be seen in our online appendix~\cite{appendix}.

\item 
In RQ$_3$, when studying the generalization of our models,
we observed that class-level refactoring models outperformed method- and variable-level refactorings. 
We took a harder look at our data and noticed that this phenomenon tends to happen when 
models are built from smaller datasets, F-Droid and Apache.  
When training our models with GitHub data (the largest dataset), the phenomenon still occurs, although with a smaller difference. Nevertheless, we can not offer a clear explanation on why that happens, 
based on the data we collected. There might be an unseen factor which we did not collect data and analyze. 
Future work should understand the reasons for this phenomenon.

\end{itemize}

\subsection{External validity} 
Threats to external validity concern the generalization of results.

\begin{itemize}
\item
Our results are based only on open source projects, 
which might affect their generalizability to industrial settings. 
It is worth mentioning, however, 
that our sample contains many industrial-scale projects that span different domains. 
To the best of our knowledge, 
this is the most extensive study of ML algorithms for the prediction of refactorings to date. 
Nevertheless, replicating this research in a large dataset of industry projects is necessary.

\item 
One of our goals was to understand whether ML models trained on a set of systems are able to accurately recommend refactoring operations to improve completely different software systems. 
We experimented with different ecosystems as an approximation for ``completely different software systems''.
While we believe this is a reasonable approximation, 
we are not able to make strong assumptions about the accuracy of those models in large-scale enterprise industrial systems. 
We suggest that researchers perform case studies together with industrial partners in hopes of providing evidence to support such hypothesis.

\item
Moreover,  
since we considered Java as the language of choice, 
we cannot be sure that our results carry over to other programming languages. 
Thus, replications of this study are needed for different programming languages. 
However, 
we cannot think of any reason why the results would be different for other imperative object-oriented languages.
\end{itemize}

\section{Conclusion}
\label{sec:conclusion}

Supervised ML algorithms are effective in predicting
refactoring opportunities 
and might indeed support developers in making faster and more educated decisions 
concerning what to refactor.

Our main findings show that:  

\begin{enumerate}
    \item Random Forest models outperform other ML models in predicting software refactoring; 
    \item Process and ownership metrics seem to play a crucial role in the creation of better models; and
    \item Models trained with data from heterogeneous projects generalize better and achieve good performance.
\end{enumerate}

More importantly, this paper shows that \textbf{ML algorithms 
can accurately model the refactoring recommendation problem}. 
We hope that this paper will pave
the way for more data-driven refactoring recommendation tools.

Given that we are more confident that ML models might provide accurate 
recommendations to developers, the next step of this research should work on devising and building the necessary tools to 
deploy and perform case studies on the efficiency of refactoring recommendation models in the wild.

\section*{Acknowledgements}

We thank Prof. Dr. Alfredo Goldman (University of São Paulo), Diogo Pina (University of São Paulo), and Matheus Flauzino (Federal University of Lavras) for their feedback on the early steps of this work.

\bibliographystyle{IEEEtranN}
\bibliography{paper}

\begin{thebibliography}{84}
\providecommand{\natexlab}[1]{#1}
\providecommand{\url}[1]{#1}
\csname url@samestyle\endcsname
\providecommand{\newblock}{\relax}
\providecommand{\bibinfo}[2]{#2}
\providecommand{\BIBentrySTDinterwordspacing}{\spaceskip=0pt\relax}
\providecommand{\BIBentryALTinterwordstretchfactor}{4}
\providecommand{\BIBentryALTinterwordspacing}{\spaceskip=\fontdimen2\font plus
\BIBentryALTinterwordstretchfactor\fontdimen3\font minus
  \fontdimen4\font\relax}
\providecommand{\BIBforeignlanguage}[2]{{%
\expandafter\ifx\csname l@#1\endcsname\relax
\typeout{** WARNING: IEEEtranN.bst: No hyphenation pattern has been}%
\typeout{** loaded for the language `#1'. Using the pattern for}%
\typeout{** the default language instead.}%
\else
\language=\csname l@#1\endcsname
\fi
#2}}
\providecommand{\BIBdecl}{\relax}
\BIBdecl

\bibitem[Fowler et~al.(1999)Fowler, Beck, Brant, Opdyke, and
  Roberts]{Fowler1999-go}
M.~Fowler, K.~Beck, J.~Brant, W.~Opdyke, and D.~Roberts, \emph{Refactoring:
  Improving the Design of Existing Code}.\hskip 1em plus 0.5em minus
  0.4em\relax Addison-Wesley, 1999.

\bibitem[AlOmar et~al.(2019)AlOmar, Mkaouer, Ouni, and Kessentini]{Eman_2019}
E.~A. AlOmar, M.~W. Mkaouer, A.~Ouni, and M.~Kessentini, ``Do design metrics
  capture developers perception of quality? an empirical study on self-affirmed
  refactoring activities,'' in \emph{Proceedings of the 13th ACM/IEEE
  International Symposium on Empirical Software Engineering and Measurement
  (ESEM 2019)}, 2019, pp. 300--311.

\bibitem[{Kataoka} et~al.(2002){Kataoka}, {Imai}, {Andou}, and
  {Fukaya}]{Kataoka_2012}
Y.~{Kataoka}, T.~{Imai}, H.~{Andou}, and T.~{Fukaya}, ``A quantitative
  evaluation of maintainability enhancement by refactoring,'' in
  \emph{International Conference on Software Maintenance, 2002. Proceedings.},
  2002, pp. 576--585.

\bibitem[Leitch and Stroulia(2004)]{leitch2004assessing}
R.~Leitch and E.~Stroulia, ``Assessing the maintainability benefits of design
  restructuring using dependency analysis,'' in \emph{Proceedings. 5th
  International Workshop on Enterprise Networking and Computing in Healthcare
  Industry (IEEE Cat. No. 03EX717)}, 2004, pp. 309--322.

\bibitem[Alshayeb(2009)]{alshayeb2009empirical}
M.~Alshayeb, ``Empirical investigation of refactoring effect on software
  quality,'' \emph{Information and software technology}, vol.~51, no.~9, pp.
  1319--1326, 2009.

\bibitem[Shatnawi and Li(2011)]{shatnawi2011empirical}
R.~Shatnawi and W.~Li, ``An empirical assessment of refactoring impact on
  software quality using a hierarchical quality model,'' \emph{International
  Journal of Software Engineering and Its Applications}, vol.~5, no.~4, pp.
  127--149, 2011.

\bibitem[Kim et~al.(2012)Kim, Zimmermann, and Nagappan]{kim2012field}
M.~Kim, T.~Zimmermann, and N.~Nagappan, ``A field study of refactoring
  challenges and benefits,'' in \emph{Proceedings of the ACM SIGSOFT 20th
  International Symposium on the Foundations of Software Engineering}, 2012,
  p.~50.

\bibitem[Kruchten et~al.(2012)Kruchten, Nord, and
  Ozkaya]{kruchten2012technical}
P.~Kruchten, R.~L. Nord, and I.~Ozkaya, ``Technical debt: From metaphor to
  theory and practice,'' \emph{IEEE Software}, vol.~29, no.~6, pp. 18--21,
  2012.

\bibitem[Beller et~al.(2016)Beller, Bholanath, McIntosh, and
  Zaidman]{beller2016analyzing}
M.~Beller, R.~Bholanath, S.~McIntosh, and A.~Zaidman, ``Analyzing the state of
  static analysis: A large-scale evaluation in open source software,'' in
  \emph{2016 IEEE 23rd International Conference on Software Analysis,
  Evolution, and Reengineering (SANER)}, vol.~1.\hskip 1em plus 0.5em minus
  0.4em\relax IEEE, 2016, pp. 470--481.

\bibitem[Ayewah et~al.(2008)Ayewah, Pugh, Hovemeyer, Morgenthaler, and
  Penix]{ayewah2008using}
N.~Ayewah, W.~Pugh, D.~Hovemeyer, J.~D. Morgenthaler, and J.~Penix, ``Using
  static analysis to find bugs,'' \emph{IEEE software}, vol.~25, no.~5, pp.
  22--29, 2008.

\bibitem[Habchi et~al.(2018)Habchi, Blanc, and Rouvoy]{habchi2018adopting}
S.~Habchi, X.~Blanc, and R.~Rouvoy, ``On adopting linters to deal with
  performance concerns in android apps,'' in \emph{Proceedings of the 33rd
  ACM/IEEE International Conference on Automated Software Engineering}, 2018,
  pp. 6--16.

\bibitem[T{\'o}masd{\'o}ttir et~al.(2018)T{\'o}masd{\'o}ttir, Aniche, and
  Van~Deursen]{tomasdottir2018adoption}
K.~F. T{\'o}masd{\'o}ttir, M.~Aniche, and A.~Van~Deursen, ``The adoption of
  javascript linters in practice: A case study on eslint,'' \emph{IEEE
  Transactions on Software Engineering}, 2018.

\bibitem[T{\'o}masd{\'o}ttir et~al.(2017)T{\'o}masd{\'o}ttir, Aniche, and van
  Deursen]{tomasdottir2017and}
K.~F. T{\'o}masd{\'o}ttir, M.~Aniche, and A.~van Deursen, ``Why and how
  javascript developers use linters,'' in \emph{2017 32nd IEEE/ACM
  International Conference on Automated Software Engineering (ASE)}.\hskip 1em
  plus 0.5em minus 0.4em\relax IEEE, 2017, pp. 578--589.

\bibitem[Johnson et~al.(2013)Johnson, Song, Murphy-Hill, and
  Bowdidge]{johnson2013don}
B.~Johnson, Y.~Song, E.~Murphy-Hill, and R.~Bowdidge, ``Why don't software
  developers use static analysis tools to find bugs?'' in \emph{Proceedings of
  the 2013 International Conference on Software Engineering}, 2013, pp.
  672--681.

\bibitem[Lanza and Marinescu(2007)]{lanza2007object}
M.~Lanza and R.~Marinescu, \emph{Object-oriented metrics in practice: using
  software metrics to characterize, evaluate, and improve the design of
  object-oriented systems}.\hskip 1em plus 0.5em minus 0.4em\relax Springer
  Science \& Business Media, 2007.

\bibitem[Moha et~al.(2009)Moha, Gueheneuc, Duchien, and Le~Meur]{moha2009decor}
N.~Moha, Y.-G. Gueheneuc, L.~Duchien, and A.-F. Le~Meur, ``Decor: A method for
  the specification and detection of code and design smells,'' \emph{IEEE
  Transactions on Software Engineering}, vol.~36, no.~1, pp. 20--36, 2009.

\bibitem[Mariani and Vergilio(2017)]{mariani2017systematic}
T.~Mariani and S.~R. Vergilio, ``A systematic review on search-based
  refactoring,'' \emph{Information and Software Technology}, vol.~83, pp.
  14--34, 2017.

\bibitem[O’Keeffe and Cinn{\'e}ide(2008)]{o2008search}
M.~O’Keeffe and M.~O. Cinn{\'e}ide, ``Search-based refactoring for software
  maintenance,'' \emph{Journal of Systems and Software}, vol.~81, no.~4, pp.
  502--516, 2008.

\bibitem[Bavota et~al.(2014)Bavota, De~Lucia, Marcus, and
  Oliveto]{bavota2014recommending}
G.~Bavota, A.~De~Lucia, A.~Marcus, and R.~Oliveto, ``Recommending refactoring
  operations in large software systems,'' in \emph{Recommendation Systems in
  Software Engineering}.\hskip 1em plus 0.5em minus 0.4em\relax Springer, 2014,
  pp. 387--419.

\bibitem[D'Ambros et~al.(2012)D'Ambros, Lanza, and Robbes]{d2012evaluating}
M.~D'Ambros, M.~Lanza, and R.~Robbes, ``Evaluating defect prediction
  approaches: a benchmark and an extensive comparison,'' \emph{Empirical
  Software Engineering}, vol.~17, no. 4-5, pp. 531--577, 2012.

\bibitem[Liu et~al.(2019)Liu, Kim, Bissyand{\'e}, Kim, Kim, Koyuncu, Kim, and
  Traon]{Liu_2019}
K.~Liu, D.~Kim, T.~F. Bissyand{\'e}, T.~Kim, K.~Kim, A.~Koyuncu, S.~Kim, and
  Y.~L. Traon, ``Learning to spot and refactor inconsistent method names,'' in
  \emph{Proceedings of the 41st International Conference on Software
  Engineering}, 2019, pp. 1--12.

\bibitem[Azeem et~al.(2019{\natexlab{a}})Azeem, Palomba, Shi, and
  Wang]{Muhammad_2019}
M.~I. Azeem, F.~Palomba, L.~Shi, and Q.~Wang, ``Machine learning techniques for
  code smell detection: A systematic literature review and meta-analysis,''
  \emph{Information and Software Technology}, vol. 108, pp. 115 -- 138, 2019.

\bibitem[Tsantalis et~al.(2018)Tsantalis, Mansouri, Eshkevari, Mazinanian, and
  Dig]{Tsantalis_2018}
N.~Tsantalis, M.~Mansouri, L.~M. Eshkevari, D.~Mazinanian, and D.~Dig,
  ``Accurate and efficient refactoring detection in commit history,'' in
  \emph{Proceedings of the 40th International Conference on Software
  Engineering}, 2018, pp. 483--494.

\bibitem[Deursen et~al.(2001)Deursen, Moonen, Bergh, and
  Kok]{van2001refactoring}
A.~V. Deursen, L.~Moonen, A.~V.~D. Bergh, and G.~Kok, ``Refactoring test
  code,'' in \emph{Proceedings of the 2nd international conference on extreme
  programming and flexible processes in software engineering (XP2001)}, 2001,
  pp. 92--95.

\bibitem[Bavota et~al.(2012)Bavota, Qusef, Oliveto, Lucia, and
  Binkley]{bavota2012empirical}
G.~Bavota, A.~Qusef, R.~Oliveto, A.~D. Lucia, and D.~Binkley, ``An empirical
  analysis of the distribution of unit test smells and their impact on software
  maintenance,'' in \emph{2012 28th IEEE International Conference on Software
  Maintenance (ICSM)}, 2012, pp. 56--65.

\bibitem[Bavota et~al.(2015)Bavota, Qusef, Oliveto, Lucia, and
  Binkley]{bavota2015test}
------, ``Are test smells really harmful? an empirical study,'' \emph{Empirical
  Software Engineering}, vol.~20, no.~4, pp. 1052--1094, 2015.

\bibitem[Scalabrino et~al.(2017)Scalabrino, Bavota, Vendome,
  Linares-V{\'a}squez, Poshyvanyk, and Oliveto]{scalabrino2017automatically}
S.~Scalabrino, G.~Bavota, C.~Vendome, M.~Linares-V{\'a}squez, D.~Poshyvanyk,
  and R.~Oliveto, ``Automatically assessing code understandability: How far are
  we?'' in \emph{Proceedings of the 32nd IEEE/ACM International Conference on
  Automated Software Engineering}, 2017, pp. 417--427.

\bibitem[McIntosh et~al.(2014)McIntosh, Kamei, Adams, and
  Hassan]{mcintosh2014impact}
S.~McIntosh, Y.~Kamei, B.~Adams, and A.~E. Hassan, ``The impact of code review
  coverage and code review participation on software quality: A case study of
  the qt, vtk, and itk projects,'' in \emph{Proceedings of the 11th Working
  Conference on Mining Software Repositories}, 2014, pp. 192--201.

\bibitem[Kaur and
  Singh(2019)]{how_does_oo_code_refactoring_influence_software_quality}
S.~Kaur and P.~Singh, ``How does object-oriented code refactoring influence
  software quality? research landscape and challenges,'' \emph{Journal of
  Systems and Software}, vol. 157, pp. 110--394, 2019.

\bibitem[Aniche et~al.()Aniche, Maziero, Durelli, and Durelli]{appendix}
M.~Aniche, E.~Maziero, R.~Durelli, and V.~Durelli, ``Appendix: The
  effectiveness of supervised machine learning algorithms in predicting
  software refactoring,'' appendix: \url{https://zenodo.org/record/3598352},
  Dataset: \url{https://zenodo.org/record/3547639}, Final models:
  \url{https://zenodo.org/record/3598361}, Source code:
  \url{http://github.com/refactoring-ai/predicting-refactoring-ml} (commit
  e397ab8f).

\bibitem[Chidamber and Kemerer(1994)]{chidamber1994metrics}
S.~R. Chidamber and C.~F. Kemerer, ``A metrics suite for object oriented
  design,'' \emph{IEEE Transactions on software engineering}, vol.~20, no.~6,
  pp. 476--493, 1994.

\bibitem[Rahman and Devanbu(2013)]{rahman2013and}
F.~Rahman and P.~Devanbu, ``How, and why, process metrics are better,'' in
  \emph{2013 35th International Conference on Software Engineering (ICSE)},
  2013, pp. 432--441.

\bibitem[Madeyski and Jureczko(2015)]{madeyski2015process}
L.~Madeyski and M.~Jureczko, ``Which process metrics can significantly improve
  defect prediction models? an empirical study,'' \emph{Software Quality
  Journal}, vol.~23, no.~3, pp. 393--422, 2015.

\bibitem[Bird et~al.(2011)Bird, Nagappan, Murphy, Gall, and
  Devanbu]{dont_touch_my_code}
C.~Bird, N.~Nagappan, B.~Murphy, H.~Gall, and P.~Devanbu, ``Don't touch my
  code!: Examining the effects of ownership on software quality,'' in
  \emph{Proceedings of the 19th ACM SIGSOFT Symposium and the 13th European
  Conference on Foundations of Software Engineering}, 2011, pp. 4--14.

\bibitem[Higo et~al.(2020)Higo, Hayashi, and Kusumoto]{higo2020tracking}
Y.~Higo, S.~Hayashi, and S.~Kusumoto, ``On tracking java methods with git
  mechanisms,'' \emph{Journal of Systems and Software}, p. 110571, 2020.

\bibitem[Pedregosa et~al.(2011)Pedregosa, Varoquaux, Gramfort, Michel, Thirion,
  Grisel, Blondel, Prettenhofer, Weiss, Dubourg, Vanderplas, Passos,
  Cournapeau, Brucher, Perrot, and Duchesnay]{scikit-learn}
F.~Pedregosa, G.~Varoquaux, A.~Gramfort, V.~Michel, B.~Thirion, O.~Grisel,
  M.~Blondel, P.~Prettenhofer, R.~Weiss, V.~Dubourg, J.~Vanderplas, A.~Passos,
  D.~Cournapeau, M.~Brucher, M.~Perrot, and E.~Duchesnay, ``Scikit-learn:
  Machine learning in {P}ython,'' \emph{Journal of Machine Learning Research},
  vol.~12, no. MISSING, pp. 2825--2830, 2011.

\bibitem[Bishop(2006)]{bishop2006_lr}
C.~M. Bishop, \emph{Pattern Recognition and Machine Learning (Information
  Science and Statistics)}.\hskip 1em plus 0.5em minus 0.4em\relax
  Springer-Verlag, 2006.

\bibitem[Zhang(2004)]{Zhang04theoptimality}
H.~Zhang, ``The optimality of naïve bayes,'' in \emph{In FLAIRS2004
  conference}, 2004.

\bibitem[Cortes and Vapnik(1995)]{cortes1995_svm}
C.~Cortes and V.~Vapnik, ``Support-vector networks,'' in \emph{Machine
  Learning}, 1995, pp. 273--297.

\bibitem[Quinlan(1993)]{quinlan1993_dt}
J.~R. Quinlan, \emph{C4.5: Programs for Machine Learning}.\hskip 1em plus 0.5em
  minus 0.4em\relax Morgan Kaufmann Publishers Inc., 1993.

\bibitem[Breiman(2001)]{breiman2001_rf}
L.~Breiman, ``Random forests,'' \emph{Machine Learning}, vol.~45, no.~1, pp.
  5--32, 2001.

\bibitem[Goodfellow et~al.(2016)Goodfellow, Bengio, and
  Courville]{Goodfellow-et-al-2016}
I.~Goodfellow, Y.~Bengio, and A.~Courville, \emph{Deep Learning}.\hskip 1em
  plus 0.5em minus 0.4em\relax MIT Press, 2016,
  http://www.deeplearningbook.org.

\bibitem[Ioffe and Szegedy(2015)]{Normalization43442}
S.~Ioffe and C.~Szegedy, ``Batch normalization: Accelerating deep network
  training by reducing internal covariate shift,'' vol.~37, pp. 448--456, 2015.

\bibitem[He and Garcia(2009)]{he2009learning}
H.~He and E.~A. Garcia, ``Learning from imbalanced data,'' \emph{IEEE
  Transactions on knowledge and data engineering}, vol.~21, no.~9, pp.
  1263--1284, 2009.

\bibitem[Rapp et~al.(2019)Rapp, Menc{\'\i}a, and
  F{\"u}rnkranz]{rapp2019simplifying}
M.~Rapp, E.~L. Menc{\'\i}a, and J.~F{\"u}rnkranz, ``{Simplifying Random
  Forests: On the Trade-off between Interpretability and Accuracy},'' in
  \emph{Proceedings of the 1st Workshop on Deep Continuous-Discrete Machine
  Learning}, 2019, pp. 1--3.

\bibitem[Kim et~al.(2018)Kim, Wattenberg, Gilmer, Cai, Wexler, Viegas, and
  sayres]{kim2017interpretability}
B.~Kim, M.~Wattenberg, J.~Gilmer, C.~Cai, J.~Wexler, F.~Viegas, and R.~sayres,
  ``{Interpretability Beyond Feature Attribution: Quantitative Testing with
  Concept Activation Vectors ({TCAV})},'' vol.~80, pp. 2668--2677, 2018.

\bibitem[Lehman et~al.(1997)Lehman, Ramil, Wernick, Perry, and
  Turski]{Lehman1997}
M.~M. Lehman, J.~F. Ramil, P.~D. Wernick, D.~E. Perry, and W.~M. Turski,
  ``Metrics and laws of software evolution - the nineties view,'' in
  \emph{Proceedings of the 4th International Symposium on Software Metrics},
  1997, pp. 20--32.

\bibitem[Lehman(1996)]{lehman1996laws}
M.~M. Lehman, ``Laws of software evolution revisited,'' in \emph{European
  Workshop on Software Process Technology}, 1996, pp. 108--124.

\bibitem[Alon et~al.(2019)Alon, Zilberstein, Levy, and Yahav]{code2vec}
\BIBentryALTinterwordspacing
U.~Alon, M.~Zilberstein, O.~Levy, and E.~Yahav, ``Code2vec: Learning
  distributed representations of code,'' \emph{Proc. ACM Program. Lang.},
  vol.~3, no. POPL, pp. 40:1--40:29, 2019. [Online]. Available:
  \url{http://doi.acm.org/10.1145/3290353}
\BIBentrySTDinterwordspacing

\bibitem[Mikolov et~al.(2013)Mikolov, Chen, Corrado, and
  Dean]{mikolov2013efficient}
T.~Mikolov, K.~Chen, G.~S. Corrado, and J.~Dean, ``{Efficient Estimation of
  Word Representations in Vector Space},'' in \emph{International Conference on
  Learning Representations}, 2013, pp. 1--12.

\bibitem[Kerievsky(2004)]{kerievsky2005refactoring}
J.~Kerievsky, \emph{{Refactoring to Patterns}}.\hskip 1em plus 0.5em minus
  0.4em\relax Addison-Wesley Professional, 2004.

\bibitem[Silva et~al.(2016)Silva, Tsantalis, and Valente]{silva2016we}
D.~Silva, N.~Tsantalis, and M.~T. Valente, ``Why we refactor? confessions of
  github contributors,'' in \emph{Proceedings of the 2016 24th ACM SIGSOFT
  International Symposium on Foundations of Software Engineering}, 2016, pp.
  858--870.

\bibitem[Allamanis et~al.(2018)Allamanis, Barr, Devanbu, and
  Sutton]{allamanis2018survey}
M.~Allamanis, E.~T. Barr, P.~Devanbu, and C.~Sutton, ``A survey of machine
  learning for big code and naturalness,'' \emph{ACM Computing Surveys (CSUR)},
  vol.~51, no.~4, p.~81, 2018.

\bibitem[Hindle et~al.(2016)Hindle, Barr, Gabel, Su, and
  Devanbu]{Hindle:2016:NS:2930840.2902362}
\BIBentryALTinterwordspacing
A.~Hindle, E.~T. Barr, M.~Gabel, Z.~Su, and P.~Devanbu, ``On the naturalness of
  software,'' \emph{Commun. ACM}, vol.~59, no.~5, pp. 122--131, Apr. 2016.
  [Online]. Available: \url{http://doi.acm.org/10.1145/2902362}
\BIBentrySTDinterwordspacing

\bibitem[Hellendoorn and Devanbu(2017)]{Hellendoorn:2017:DNN:3106237.3106290}
\BIBentryALTinterwordspacing
V.~J. Hellendoorn and P.~Devanbu, ``Are deep neural networks the best choice
  for modeling source code?'' in \emph{Proceedings of the 2017 11th Joint
  Meeting on Foundations of Software Engineering}, ser. ESEC/FSE 2017.\hskip
  1em plus 0.5em minus 0.4em\relax New York, NY, USA: ACM, 2017, pp. 763--773.
  [Online]. Available: \url{http://doi.acm.org/10.1145/3106237.3106290}
\BIBentrySTDinterwordspacing

\bibitem[Allamanis et~al.(2015)Allamanis, Barr, Bird, and
  Sutton]{Allamanis_2015}
M.~Allamanis, E.~T. Barr, C.~Bird, and C.~Sutton, ``Suggesting accurate method
  and class names,'' in \emph{Proceedings of the 2015 10th Joint Meeting on
  Foundations of Software Engineering}, ser. ESEC/FSE 2015, 2015, pp. 38--49.

\bibitem[Wong et~al.(2013)Wong, Yang, and Tan]{Wong_2013}
E.~Wong, J.~Yang, and L.~Tan, ``Autocomment: Mining question and answer sites
  for automatic comment generation,'' in \emph{2013 28th IEEE/ACM International
  Conference on Automated Software Engineering (ASE)}, 2013, pp. 562--567.

\bibitem[Jiang et~al.(2017)Jiang, Armaly, and McMillan]{jiang2017automatically}
S.~Jiang, A.~Armaly, and C.~McMillan, ``Automatically generating commit
  messages from diffs using neural machine translation,'' in \emph{Proceedings
  of the 32nd IEEE/ACM International Conference on Automated Software
  Engineering}, 2017, pp. 135--146.

\bibitem[{Deep Singh} and {Chug}(2017)]{Deep_2017}
P.~{Deep Singh} and A.~{Chug}, ``Software defect prediction analysis using
  machine learning algorithms,'' in \emph{2017 7th International Conference on
  Cloud Computing, Data Science Engineering - Confluence}, 2017.

\bibitem[Sutskever et~al.(2014)Sutskever, Vinyals, and
  Le]{Sutskever2014SequenceTS}
I.~Sutskever, O.~Vinyals, and Q.~V. Le, ``Sequence to sequence learning with
  neural networks,'' in \emph{NIPS}, 2014.

\bibitem[Buciluǎ et~al.(2006)Buciluǎ, Caruana, and
  Niculescu-Mizil]{bucilua2006model}
C.~Buciluǎ, R.~Caruana, and A.~Niculescu-Mizil, ``Model compression,'' in
  \emph{Proceedings of the 12th ACM SIGKDD international conference on
  Knowledge discovery and data mining}, 2006, pp. 535--541.

\bibitem[Kondo et~al.(2019)Kondo, Bezemer, Kamei, Hassan, and
  Mizuno]{kondo2019impact}
M.~Kondo, C.-P. Bezemer, Y.~Kamei, A.~E. Hassan, and O.~Mizuno, ``The impact of
  feature reduction techniques on defect prediction models,'' \emph{Empirical
  Software Engineering}, pp. 1--39, 2019.

\bibitem[Sadowski et~al.(2015)Sadowski, Gogh, Jaspan, S{\"o}derberg, and
  Winter]{sadowski2015tricorder}
C.~Sadowski, J.~V. Gogh, C.~Jaspan, E.~S{\"o}derberg, and C.~Winter,
  ``Tricorder: Building a program analysis ecosystem,'' in \emph{Proceedings of
  the 37th International Conference on Software Engineering-Volume 1}, 2015,
  pp. 598--608.

\bibitem[Sadowski et~al.(2018)Sadowski, Aftandilian, Eagle, Miller-Cushon, and
  Jaspan]{sadowski2018lessons}
C.~Sadowski, E.~Aftandilian, A.~Eagle, L.~Miller-Cushon, and C.~Jaspan,
  ``Lessons from building static analysis tools at google,''
  \emph{Communications of the ACM}, vol.~61, no.~4, pp. 58--66, 2018.

\bibitem[Kumar et~al.(2019)Kumar, Bansal, Maddila, Sharma, Martelock, and
  Bhargava]{kumar2019building}
R.~Kumar, C.~Bansal, C.~Maddila, N.~Sharma, S.~Martelock, and R.~Bhargava,
  ``Building sankie: an ai platform for devops,'' in \emph{2019 IEEE/ACM 1st
  International Workshop on Bots in Software Engineering (BotSE)}.\hskip 1em
  plus 0.5em minus 0.4em\relax IEEE, 2019, pp. 48--53.

\bibitem[Goldberg and Levy(2014)]{goldberg2014word2vec}
Y.~Goldberg and O.~Levy, ``word2vec explained: deriving mikolov et al.'s
  negative-sampling word-embedding method,'' 2014.

\bibitem[Devlin et~al.(2018)Devlin, Chang, Lee, and Toutanova]{devlin2018bert}
J.~Devlin, M.-W. Chang, K.~Lee, and K.~Toutanova, ``Bert: Pre-training of deep
  bidirectional transformers for language understanding,'' 2018.

\bibitem[{Mens} and {Tourw\'e}(2004)]{a_survey_of_software_refactoring}
T.~{Mens} and T.~{Tourw\'e}, ``A survey of software refactoring,'' \emph{IEEE
  Transactions on Software Engineering}, vol.~30, no.~2, pp. 126--139, 2004.

\bibitem[Baqais and Alshayeb(2019)]{automatic_software_refactoring_a_slr}
A.~A.~B. Baqais and M.~Alshayeb, ``Automatic software refactoring: a systematic
  literature review,'' \emph{Software Quality Journal}, 2019.

\bibitem[Zhang et~al.(2011)Zhang, Hall, and
  Baddoo]{code_bad_smells_a_review_of_current_knowledge}
M.~Zhang, T.~Hall, and N.~Baddoo, ``Code bad smells: A review of current
  knowledge,'' \emph{Journal of Software Maintenance and Evolution: Research
  and Practice}, vol.~23, no.~3, pp. 179--202, 2011.

\bibitem[{Al
  Dallal}(2015)]{identifying_refactoring_opportunities_in_oo_code_slr}
J.~{Al Dallal}, ``Identifying refactoring opportunities in object-oriented
  code: A systematic literature review,'' \emph{Information and Software
  Technology}, vol.~58, pp. 231--249, 2015.

\bibitem[Singh and
  Kaur(2018)]{refactoring_for_disclosing_code_smells_in_oo_software}
S.~Singh and S.~Kaur, ``A systematic literature review: Refactoring for
  disclosing code smells in object oriented software,'' \emph{Ain Shams
  Engineering Journal}, vol.~9, no.~4, pp. 2129--2151, 2018.

\bibitem[Santos et~al.(2018)Santos, Rocha-Junior, Prates, do~Nascimento,
  Freitas, and de~Mendon{\c c}a]{a_systematic_review_on_the_code_smell_effect}
J.~A.~M. Santos, J.~B. Rocha-Junior, L.~C.~L. Prates, R.~S. do~Nascimento,
  M.~F. Freitas, and M.~G. de~Mendon{\c c}a, ``A systematic review on the code
  smell effect,'' \emph{Journal of Systems and Software}, vol. 144, pp.
  450--477, 2018.

\bibitem[Fernandes et~al.(2016)Fernandes, Oliveira, Vale, Paiva, and
  Figueiredo]{a_review_based_comparative_study_of_bad_smell_detection_tools}
E.~Fernandes, J.~Oliveira, G.~Vale, T.~Paiva, and E.~Figueiredo, ``A
  review-based comparative study of bad smell detection tools,'' in
  \emph{Proceedings of the 20th International Conference on Evaluation and
  Assessment in Software Engineering}.\hskip 1em plus 0.5em minus 0.4em\relax
  ACM, 2016, pp. 18:1--18:12.

\bibitem[Rasool and Arshad(2015)]{a_review_of_code_smell_mining_techniques}
G.~Rasool and Z.~Arshad, ``A review of code smell mining techniques,''
  \emph{Journal of Software: Evolution and Process}, vol.~27, no.~11, pp.
  867--895, 2015.

\bibitem[Mohan and
  Greer(2018)]{a_survey_of_search_based_refactoring_for_software_maintenance}
M.~Mohan and D.~Greer, ``A survey of search-based refactoring for software
  maintenance,'' \emph{Journal of Software Engineering Research and
  Development}, vol.~6, no.~1, pp. 3--55, 2018.

\bibitem[Azeem et~al.(2019{\natexlab{b}})Azeem, Palomba, Shi, and
  Wang]{machine_learning_techniques_for_code_smell_detection}
M.~I. Azeem, F.~Palomba, L.~Shi, and Q.~Wang, ``Machine learning techniques for
  code smell detection: A systematic literature review and meta-analysis,''
  \emph{Information and Software Technology}, vol. 108, pp. 115--138, 2019.

\bibitem[Zhang and Mani(2003)]{mani2003knn}
J.~Zhang and I.~Mani, ``{kNN Approach to Unbalanced Data Distributions: A Case
  Study Involving Information Extraction},'' in \emph{{Proceedings of Workshop
  on Learning from Imbalanced Datasets}}, vol. 126, 2003, pp. 1--7.

\bibitem[Marinescu(2004)]{marinescu2004detection}
R.~Marinescu, ``Detection strategies: Metrics-based rules for detecting design
  flaws,'' in \emph{20th IEEE International Conference on Software Maintenance,
  2004. Proceedings.}\hskip 1em plus 0.5em minus 0.4em\relax IEEE, 2004, pp.
  350--359.

\bibitem[{Goularte Carvalho} et~al.(2019){Goularte Carvalho}, Aniche,
  Ver{\'i}ssimo, Durelli, and Gerosa]{aniche-1}
S.~{Goularte Carvalho}, M.~Aniche, J.~Ver{\'i}ssimo, R.~Durelli, and M.~Gerosa,
  ``\BIBforeignlanguage{English}{An empirical catalog of code smells for the
  presentation layer of android apps},''
  \emph{\BIBforeignlanguage{English}{Empirical Software Engineering}}, vol.~24,
  no.~6, p. 3546–3586, 2019.

\bibitem[Aniche et~al.(2018)Aniche, Bavota, Treude, Gerosa, and {van
  Deursen}]{aniche-2}
M.~Aniche, G.~Bavota, C.~Treude, M.~Gerosa, and A.~{van Deursen},
  ``\BIBforeignlanguage{English}{Code smells for model-view-controller
  architectures},'' \emph{\BIBforeignlanguage{English}{Empirical Software
  Engineering}}, vol.~23, no.~4, pp. 2121--2157, 2018.

\bibitem[Aniche et~al.(2016{\natexlab{a}})Aniche, Bavota, Treude, {van
  Deursen}, and Gerosa]{aniche-3}
M.~Aniche, G.~Bavota, C.~Treude, A.~{van Deursen}, and M.~Gerosa,
  ``\BIBforeignlanguage{English}{A validated set of smells in
  model-view-controller architectures},'' in
  \emph{\BIBforeignlanguage{English}{Proceedings 2016 IEEE International
  Conference on Software Maintenance and Evolution, ICSME 2016}}.\hskip 1em
  plus 0.5em minus 0.4em\relax United States: IEEE, 2016, pp. 233--243.

\bibitem[Aniche et~al.(2016{\natexlab{b}})Aniche, Treude, Zaidman, {van
  Deursen}, and Gerosa]{aniche-4}
M.~Aniche, C.~Treude, A.~Zaidman, A.~{van Deursen}, and M.~Gerosa,
  ``\BIBforeignlanguage{English}{Satt: Tailoring code metric thresholds for
  different software architectures},'' in
  \emph{\BIBforeignlanguage{English}{16th International Working Conference on
  Source Code Analysis and Manipulation (SCAM)}}.\hskip 1em plus 0.5em minus
  0.4em\relax United States: IEEE, 2016, pp. 41--50.

\bibitem[Pascarella et~al.(2019)Pascarella, Palomba, and
  Bacchelli]{pascarella2019fine}
L.~Pascarella, F.~Palomba, and A.~Bacchelli, ``Fine-grained just-in-time defect
  prediction,'' \emph{Journal of Systems and Software}, vol. 150, pp. 22--36,
  2019.

\end{thebibliography}
\end{document}